\documentstyle[12pt,fleqn,epsf,epsfig,axodraw]{article}
\renewcommand{\theequation}{\arabic{equation}}

\def\lsim{\raise0.3ex\hbox{$\;<$\kern-0.75em\raise-1.1ex\hbox{$\sim\;$}}}
\def\gsim{\raise0.3ex\hbox{$\;>$\kern-0.75em\raise-1.1ex\hbox{$\sim\;$}}}

\begin{document}
\setlength{\unitlength}{1cm}
\setlength{\mathindent}{0cm}
\thispagestyle{empty}

\begin{flushright}
	WUE-ITP-04-032\\
	UWThPh-2004-25\\
	HEPHY-PUB 796/04\\
	hep-ph/0410054
\end{flushright}

\vskip .8cm
\begin{center}
	{\Large \bf 
		CP sensitive observables in chargino production 
		and decay into a $W$ boson
		}
\vskip 2.5em
{\large
{\sc  O.~Kittel$^{a}$\footnote{e-mail:
		  kittel@physik.uni-wuerzburg.de},
		A.~Bartl$^{b}$\footnote{e-mail:
        bartl@ap.univie.ac.at},
      H.~Fraas$^{a}$\footnote{e-mail:
        fraas@physik.uni-wuerzburg.de},
		W.~Majerotto$^{c}$\footnote{e-mail:
        majer@qhepu3.oeaw.ac.at}
}}\\[1ex]
{\normalsize \it
$^{a}$ Institut f\"ur Theoretische Physik, Universit\"at
W\"urzburg, Am Hubland, D-97074~W\"urzburg, Germany}\\
{\normalsize \it
$^{b}$ Institut f\"ur Theoretische Physik, Universit\"at Wien, 
Boltzmanngasse 5, A-1090 Wien, Austria}\\
{\normalsize \it
$^{c}$ Institut f\"ur Hochenergiephysik, \"Osterreichische
Akademie der Wissenschaften, Nikolsdorfergasse 18, 
A-1050 Wien, Austria}\\
\vskip 1em
\end{center} \par
\vskip .8cm

\begin{abstract}

We study CP sensitive observables in chargino production 
in electron-positron collisions
with subsequent two-body decay of one chargino
into a $W$ boson.
We identify the CP odd  elements of the $W$ boson
density matrix and  propose CP sensitive 
triple-product asymmetries of the chargino decay products. 
We calculate the density-matrix elements, the CP asymmetries 
and the cross sections 
in the Minimal Supersymmetric Standard Model with complex parameters 
$\mu$ and $M_1$ for  an $e^+e^-$ linear collider with 
$\sqrt{s}=800$ GeV and longitudinally polarized beams. 
The asymmetries can reach  $7\%$ 
and we discuss the feasibility of measuring these asymmetries.

\end{abstract}

\newpage

\section{Introduction}

In the Minimal Supersymmetric Standard Model (MSSM) \cite{haberkane} 
several supersymmetric (SUSY) parameters  can be complex. 
In the chargino sector of the MSSM this is 
the Higgsino mass parameter $\mu = |\mu| \, e^{ i\,\varphi_{\mu}}$, 
and in the neutralino sector of the MSSM also the 
$U(1)$ gaugino mass parameter $M_1 = |M_1| \, e^{ i\,\varphi_{M_1}}$
can have a physical phase \cite{Dugan}. 
Usually it is claimed that these phases, 
in particular $\varphi_{\mu}$, have to be small \cite{nath,edms},
due to the experimental upper 
bounds of the electric dipole moments (EDMs) 
of electron and neutron.
For example in the constrained MSSM
$|\varphi_{\mu}|$
has to be smaller than
$\pi/10$ \cite{edms} for a 
supersymmetric (SUSY) particle spectrum of the order a few 100 GeV.
However, 
	the EDM restrictions may be less stringent if cancellations among 
	the different SUSY contributions occur \cite{nath}.
	The restrictions may disappear if also lepton flavor violating 
	terms are included \cite{BMPW}.
Thus, the restrictions on the phases  are very model dependent 
and independent measurements are desirable.
The study of chargino production
at an $e^+e^-$ linear collider \cite{LC} will play an important role.
By measurements of the chargino masses and cross sections,
a method has been developed in \cite{choi1, choigaiss}
to determine $\cos\varphi_{\mu}$,
in addition to the parameters $M_2$, $|\mu|$ and $\tan\beta$. 
However, also the sign of $\varphi_{\mu}$ has to be determined 
unambiguously by using CP sensitive observables.
One such observable is the chargino polarization 
perpendicular to the production plane \cite{choi1,choigaiss}. 
At tree level, this polarization  leads to triple-product asymmetries 
\cite{tripleprods,karl,olaf1,olafz}.
For chargino production and subsequent two-body decay of one chargino 
into a sneutrino, such an asymmetry can be 
as large as 30\%
\cite{olaf2} and it will allow us to constrain $\varphi_{\mu}$. 
In the present work we will study  chargino production and decay into
a $W$ boson. We will show that, due to the spin correlations
between the chargino and the $W$ boson, also an asymmetry is obtained 
which is sensitive to $\varphi_{M_1}$.

We study chargino production
\begin{eqnarray} \label{production}
	e^++e^-&\to&\tilde\chi^+_i(p_{\chi^+_i}, \lambda_i)+
	            \tilde\chi^-_j(p_{\chi^-_j}, \lambda_j), 
\end{eqnarray}
with longitudinally polarized beams and the subsequent two-body decay
\begin{eqnarray} \label{decay_1}
	\tilde\chi^+_i &\to&\tilde \chi^0_n(p_{\chi^0_n}, \lambda_n) 
	+W^+(p_W, \lambda_k), 
\end{eqnarray}
where $p$ and $\lambda$ denote momentum and helicity.
We define the triple product
 \begin{eqnarray}\label{tripleproduct1}
	 {\mathcal T}_{I} &=& 
	 {\mathbf p}_{e^-}\cdot({\mathbf p}_{\chi^+_i} \times {\mathbf p}_W)
 \end{eqnarray}
and the T odd asymmetry
\begin{eqnarray}\label{AT1}
	 {\mathcal A}_{I}^{\rm T} &=& 
	 \frac{\sigma({\mathcal T}_{I}>0)-\sigma({\mathcal T}_{I}<0)}
	{\sigma({\mathcal T}_{I}>0)+\sigma({\mathcal T}_{I}<0)},
\end{eqnarray}
with $\sigma$ the cross section of chargino production
(\ref{production}) and decay (\ref{decay_1}).
The  asymmetry ${\mathcal A}_{I}^{\rm T}$ is sensitive
to the CP violating phase $\varphi_{\mu}$. 
In this context it is interesting to note that 
asymmetries vanish if they correspond to
a triple product which contains a  transverse
polarization vector of the $e^+$ and $e^-$ beams
\cite{choi1,holger}.

In order to probe also the phase $\varphi_{M_1}$, which enters
in the chargino decay process (\ref{decay_1}), 
we consider the subsequent hadronic decay of the $W$ boson
	\begin{eqnarray} \label{decay_2B}
	W^+ \to  c + \bar s.
\end{eqnarray}
The correlations between the $\tilde\chi^+_i$ 
polarization and the $W$ boson polarization lead 
to CP sensitive elements of the $W$ boson density matrix,
which we will identify and discuss in detail.
With the triple product
 \begin{eqnarray}\label{tripleproduct2}
	 {\mathcal T}_{II} &=& 
	 {\mathbf p}_{e^-}\cdot({\mathbf p}_{c} \times {\mathbf p}_{ \bar s}),
\end{eqnarray}
which includes the momenta of the $W$ decay products and thus
probes the $W$ polarization, we define a second T odd asymmetry
\begin{eqnarray}\label{AT2}
	 {\mathcal A}_{II}^{\rm T} &=& 
	 \frac{\sigma({\mathcal T}_{II}>0)-\sigma({\mathcal T}_{II}<0)}
	{\sigma({\mathcal T}_{II}>0)+\sigma({\mathcal T}_{II}<0)}.
\end{eqnarray}
Here, $\sigma$ is  the cross section of production (\ref{production}) 
and decay of the chargino (\ref{decay_1}) followed by that of the $W$ boson 
(\ref{decay_2B}). Owing to the spin correlations, 
${\mathcal A}_{II}^{\rm T}$ has CP sensitive
contributions from $\varphi_{\mu}$ due to the chargino production 
process (\ref{production}) and contributions due to $\varphi_{\mu}$ 
and $\varphi_{M_1}$ from the chargino decay process (\ref{decay_1}). 
We treat the decay (\ref{decay_2B}) as Standard Model process.

The T odd asymmetries ${\mathcal A}_{I}^{\rm T}$ and 
${\mathcal A}_{II}^{\rm T}$ have also absorptive contributions
from s-channel resonances or final-state interactions,
which do not signal CP violation.
In order to eliminate these contributions, 
we study the two CP odd asymmetries 
\begin{equation}
{\mathcal A}_{I} = 
\frac{1}{2}({\mathcal A}_{I}^{\rm T}-\bar{\mathcal A}_{I}^{\rm T}),
\quad 
{\mathcal A}_{II} = 
\frac{1}{2}({\mathcal A}_{II}^{\rm T}-\bar{\mathcal A}_{II}^{\rm T}),
\label{ACP}
\end{equation}
where $\bar{\mathcal A}_{I,II}^{\rm T}$ are the CP conjugated asymmetries 
for the processes 
$e^+e^-\to\tilde\chi^-_i\tilde\chi^+_j;
\tilde\chi^-_i \to W^- \tilde\chi^0_n$
and
$e^+e^-\to\tilde\chi^-_i\tilde\chi^+_j;
\tilde\chi^-_i \to W^- \tilde\chi^0_n;W^- \to\bar c s$,
respectively.

In Section 
\ref{Definitions and Formalism} 
we give our definitions and formalism used, and obtain the analytical  
formulae for the differential cross section and the $W$ boson density 
matrix. In Section \ref{T odd asymmetries} we discuss general
properties of the asymmetries. We present numerical results  
in Section \ref{Numerical results} and 
Section \ref{Summary and conclusion} gives a summary 
and conclusions.

\section{Definitions and formalism
  \label{Definitions and Formalism}}

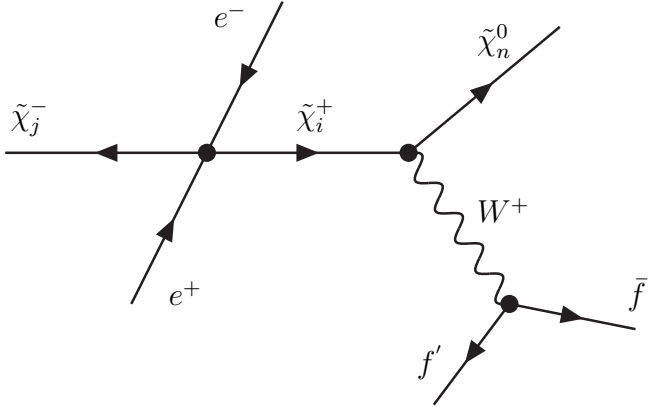
\begin{figure}[t]
\begin{picture}(5,6.)(-2,.5)
		\put(1,4.7){$\tilde\chi^-_j$}
   \put(3.7,6){$e^-$}
   \put(3.1,2.3){$e^+$}
   \put(4.8,4.7){$\tilde\chi^+_i$}
   \put(7.2,5.7){$ \tilde\chi^0_n$}
   \put(7.2,3.4){$ W^+$}
   \put(9.2,2.3){$\bar f$}
   \put(6.4,1.4){$ f^{'}$}
\end{picture}
\scalebox{1.9}{
\begin{picture}(0,0)(1.3,-0.25)
\ArrowLine(40,50)(0,50)
\Vertex(40,50){1.8}
\ArrowLine(55,80)(40,50)
\ArrowLine(25,20)(40,50)
\ArrowLine(40,50)(80,50)
\ArrowLine(80,50)(110,75)
\Photon(80,50)(100,20){2}{5}
\Vertex(80,50){1.8}
\ArrowLine(100,20)(125,15)
\ArrowLine(100,20)(85,0)
\Vertex(100,20){1.8}
\end{picture}}
\caption{\label{shematic picture}
          Schematic picture of the chargino production
          and decay process.}
\end{figure}

We give the analytical formulae for the differential cross section
of chargino production (\ref{production})
with longitudinally polarized beams and the subsequent decay chain
of one of the charginos (\ref{decay_1})
followed by the decay of the $W$ boson
 \begin{eqnarray}\label{decay_2}
	 W^+ &\to& f^{'} \bar f,
 \end{eqnarray}
which may be leptonic, $ f^{'}=\nu_{\ell},\bar f=\bar \ell$ 
with $\ell =e,\mu,\tau$ 
or hadronic, $ f^{'}=q_u, \bar f=\bar q_d$ with $q_u =u,d$ and $q_d=c,s$.
For a schematic picture of the chargino production and decay process 
see Fig.~\ref{shematic picture}.
In the following we will derive the $W$ boson spin-density matrix and 
relate it to the asymmetries ${\mathcal A}_{I}^{\rm T}$ (\ref{AT1})
and ${\mathcal A}_{II}^{\rm T}$ (\ref{AT2}).

\subsection{Lagrangian and couplings
     \label{Lagrangian}}

The MSSM interaction Lagrangians relevant for our study are 
\cite{haberkane,gudi1}:
\begin{eqnarray}
{\cal L}_{Z^0 \ell \bar \ell} &=&
- \frac{g}{\cos\theta_W} Z_{\mu}\bar \ell\gamma^{\mu}[L_{\ell}P_L+
	R_{\ell} P_R]\ell,\\
{\cal L}_{\gamma \tilde\chi^+_j \tilde\chi_i^-} &=&
- e A_{\mu} \bar{\tilde\chi}^+_i \gamma^{\mu} \tilde\chi_j^+
\delta_{ij},\quad e>0, \\
{\cal L}_{Z^0\tilde\chi_j^+\tilde\chi_i^-} &=&
 \frac{g}{\cos\theta_W}Z_{\mu}\bar{\tilde\chi}^+_i\gamma^{\mu}
[O_{ij}^{'L} P_L+O_{ij}^{'R} P_R]\tilde\chi_j^+,\\
{\cal L}_{\ell \tilde\nu\tilde\chi^+_i} &=&
- g U_{i1}^{*} \bar{\tilde\chi}^+_i P_{L} \nu \tilde\ell_L^*
- g V_{i1}^{*} \bar{\tilde\chi}_i^{+C} P_L \ell 
 \tilde{\nu}^{*}+\mbox{h.c.},\\
{\cal L}_{W^{-}\tilde\chi^+_i\tilde{\chi}^{0}_k}&=&
 g W_{\mu}^{-}\bar{\tilde{\chi}}^{0}_k\gamma^{\mu}
[O_{ki}^L P_L+O_{ki}^R P_R]\tilde\chi^+_i +
\mbox{h.c.},
\end{eqnarray}
with the couplings:
\begin{eqnarray}
 L_{\ell}&=&T_{3\ell}-e_{\ell}\sin^2\theta_W, \quad
 R_{\ell}\;=\;-e_{\ell}\sin^2\theta_W,\\
 O_{ij}^{'L}&=&-V_{i1} V_{j1}^{*}-\frac{1}{2} V_{i2} V_{j2}^{*}+
\delta_{ij} \sin^2\theta_W,\\
 O_{ij}^{'R}&=&-U_{i1}^{*} U_{j1}-\frac{1}{2} U_{i2}^{*} U_{j2}+
\delta_{ij} \sin^2\theta_W,\\
 O_{ki}^L&=&-1/\sqrt{2}\Big( \cos\beta N_{k4}-\sin\beta N_{k3}
\Big)V_{i2}^{*}
+\Big( \sin\theta_W N_{k1}+\cos\theta_W N_{k2} \Big) V_{i1}^{*},\\
O_{ki}^R&=&+1/\sqrt{2}\Big( \sin\beta N^{*}_{k4}+\cos\beta
	N^{*}_{k3}\Big) U_{i2}
+\Big( \sin\theta_W N^{*}_{k1}+\cos\theta_W N^{*}_{k2} \Big) U_{i1},
\end{eqnarray}
with $i,j=1,2$ and $k=1,\ldots,4$.
Here 
$P_{L, R}=\frac{1}{2}(1\mp \gamma_5)$, $g$ is the weak coupling
constant ($g=e/\sin\theta_W$), and $e_\ell$ and 
$T_{3 \ell}$ denote the
charge and the third component of the weak isospin of the 
lepton $\ell$. Furthermore, $\tan\beta=\frac{v_2}{v_1}$ where 
$v_{1,2}$ are the vacuum expectation values of the two 
neutral Higgs fields. The chargino-mass eigenstates 
$\tilde\chi^+_i={\chi_i^+ \choose \bar\chi_i^-}$
are defined by $\chi^{+}_i=V_{i1}w^{+}+V_{i2} h^{+}$ and 
$\chi_j^-=U_{j1}w^{-}+U_{j2} h^{-}$ with $w^{\pm}$ and $h^{\pm}$
the two-component spinor fields of the W-ino and the charged
Higgsinos, respectively. The complex unitary $2\times 2$ matrices
$U_{mn}$ and $V_{mn}$ diagonalize the chargino mass 
matrix $X_{\alpha\beta}$, $U_{m \alpha}^* X_{\alpha\beta}V_{\beta
n}^{-1}= m_{\tilde{\chi}^+_n}\delta_{mn}$, with $ m_{\tilde{\chi}^+_n}>0$. 
The complex unitary $4\times 4$ matrix $N_{ij}$
diagonalizes the neutral gaugino-Higgsino mass matrix $Y_{\alpha\beta}$, 
$N_{i \alpha}^*Y_{\alpha\beta}N_{\beta k}^{\dagger}=
m_{\tilde\chi^0_i}\delta_{ik}$, with $ m_{\tilde\chi^0_i}>0$,
in the  neutralino basis 
$\tilde{\gamma}, \tilde{Z}, \tilde{H}^0_a, \tilde{H}^0_b$.

\subsection{Helicity amplitudes
     \label{Helicity amplitudes}}
 
The helicity amplitudes
$T_P^{\lambda_i \lambda_j}$
for the production process are given in \cite{gudi1}.
Those for the chargino decay (\ref{decay_1}) are
\begin{eqnarray}
	T_{D_1,\lambda_i}^{\lambda_n\lambda_k} &=& 
	ig\bar u(p_{\chi^0_n},\lambda_n)
	\gamma^{\mu}[O_{ni}^{L}P_L + O_{ni}^{R}P_R]
		u(p_{\chi^+_i}, \lambda_i)
	\varepsilon_{\mu}^{\lambda_k\ast}
\end{eqnarray}
and those for the $W$ decay (\ref{decay_2}) are
\begin{eqnarray}
	T_{D_2,\lambda_k}^{\lambda_{f^{'}} \lambda_{\bar f}} &=& 
	i\frac{g}{\sqrt{2}}\bar u(p_{f^{'}},\lambda_{f^{'}})
	\gamma^{\mu} P_L v(p_{\bar f},\lambda_{\bar f})
	\varepsilon_{\mu}^{\lambda_k}.
\end{eqnarray}
The $W$ polarization vectors 
$\varepsilon_{\mu}^{\lambda_k},\lambda_k =0,\pm1$, are given in 
Appendix~\ref{Representation of momentum and spin vectors}.
The amplitude for the whole process 
(\ref{production}), (\ref{decay_1}), (\ref{decay_2}) is
\begin{eqnarray}
T &=& \Delta(\tilde\chi^+_i) \Delta(W^+)
\sum_{\lambda_i, \lambda_k}
T_P^{\lambda_i \lambda_j}T_{D_1,\lambda_i}^{\lambda_n\lambda_k}
T_{D_2,\lambda_k}^{\lambda_{f^{'}} \lambda_{\bar f}},
\end{eqnarray}
with the chargino propagator 
$ \Delta(\tilde{\chi}^+_i)=i/[p_{\chi^+_i}^2-m_{\chi^+_i}^2
	+im_{\chi^+_i}\Gamma_{\chi^+_i}]$ 
and the $W$ boson propagator 
$ \Delta(W^+)=i/[p_W^2-m_W^2 +im_W \Gamma_W]$.

\subsection{Cross section
     \label{Cross section}}

For the calculation of the cross section for the
combined process of chargino production (\ref{production})
and the subsequent two-body decays
(\ref{decay_1}), (\ref{decay_2}) of $\tilde\chi^+_i$
we use the same spin-density matrix formalism as in \cite{gudi1,spin}.
The (unnormalized)  spin-density matrix of the $W$ boson 
\begin{eqnarray}       \label{Wdensitymatrix}
\rho_{P}(W^+)^{\lambda_k\lambda'_k}&=&
|\Delta(\tilde\chi^+_i)|^2~
\sum_{\lambda_i,\lambda'_i}~
\rho_P   (\tilde\chi^+_i)^{\lambda_i \lambda_i'}\;
\rho_{D_1}(\tilde\chi^+_i)_{\lambda_i'\lambda_i}^{\lambda_k\lambda'_k},
\end{eqnarray}
is composed of the spin-density production matrix
\begin{eqnarray} 
	\rho_P(\tilde\chi^+_i)^{\lambda_i \lambda_i'}&=&\sum_{\lambda_j}
	T_P^{\lambda_i \lambda_j}T_P^{\lambda_i' \lambda_j \ast}
\end{eqnarray}
and the decay matrix of the chargino
\begin{eqnarray}
\rho_{D_1}(\tilde\chi^+_i)_{\lambda_i' \lambda_i}^{\lambda_k\lambda'_k} &=&
\sum_{\lambda_n}T_{D_1,\lambda_i}^{\lambda_n\lambda_k}
T_{D_1,\lambda_i'}^{\lambda_n\lambda_k'\ast}.
\end{eqnarray}
With the decay matrix for the $W$ decay
\begin{eqnarray}
	\rho_{D_2}(W^+)_{\lambda_k' \lambda_k}&=&
	\sum_{\lambda_{f^{'}}, \lambda_{\bar f}}
	T_{D_2,\lambda_k}^{\lambda_{f^{'}} \lambda_{\bar f} }
	T_{D_2,\lambda_k'}^{\lambda_{f^{'}} \lambda_{\bar f} \ast}
\end{eqnarray}
the amplitude squared for the complete process
$ e^+e^-\to\tilde\chi^+_i\tilde\chi^-_j$;
$\tilde\chi^+_i\to W^+\tilde\chi^0_n $;
$W^+ \to f^{'} \bar f$ 
can now be written
\begin{eqnarray}       \label{amplitude}
|T|^2&=&|\Delta(W^+)|^2
	\sum_{\lambda_k,\lambda'_k}~
	\rho_{P}(W^+)^{\lambda_k\lambda'_k}\;
	\rho_{D_2}(W^+)_{\lambda'_k\lambda_k}.
\end{eqnarray}
The differential cross section is then given by 
\begin{equation}\label{crossection}
	d\sigma=\frac{1}{2 s}|T|^2 
	d{\rm Lips}(s,p_{\chi^-_j },p_{\chi^0_n},p_{f^{'}},p_{\bar f}),
\end{equation}
where 
$d{\rm Lips}(s,p_{\chi^-_j },p_{\chi^0_n},p_{f^{'}},p_{\bar f})$
is the Lorentz invariant phase-space element,
see (\ref{Lips}) of Appendix \ref{Phase space}.
More details concerning kinematics and phase space  
can be found in Appendices \ref{Representation of momentum and spin vectors}
and \ref{Phase space}.

For the polarization of the decaying chargino $ \tilde \chi^+_i$
with momentum $p_{\chi^+_i}$ we introduce three space-like spin vectors
$s^{a}_{\chi^+_i}$, $a=1,2,3$, which together with 
$p_{\chi^+_i}/m_{\chi^+_i}$
form an orthonormal set  with 
$s^a_{\chi^+_i}\cdot s^b_{\chi^+_i}=-\delta^{ab}$, 
$s^a_{\chi^+_i}\cdot p_{\chi^+_i}=0$.
Then the  (unnormalized) chargino density matrix can be expanded 
in terms of the Pauli matrices $\sigma^a$, $a=1,2,3$:
\begin{eqnarray} \label{rhoP}
  \rho_P(\tilde\chi^+_i)^{\lambda_i \lambda_i'} &=&
      2(\delta_{\lambda_i \lambda_i'} P + 
       \sigma^{a}_{\lambda_i \lambda_i'}
		\Sigma_P^a),   
\end{eqnarray}
where we sum over $a$.
With our choice of the spin vectors $s^{a}_{\chi^+_i}$,
given in Appendix~\ref{Representation of momentum and spin vectors},
$\Sigma^{3}_P/P$
is the longitudinal polarization of chargino $ \tilde \chi^+_i$,
$\Sigma^{1}_P/P$ is the transverse polarization in the 
production plane and $\Sigma^{2}_P/P$ is the polarization
perpendicular to the production plane.
We give in Appendix  \ref{Chargino production matrices} the analytical 
formulae for $P$ and $\Sigma^{a}_P$ in the laboratory system.
To describe the polarization states of the $W$ boson,
we introduce  a set of spin vectors $t^c_W$, $c=1,2,3$, and choose
polarization vectors $\varepsilon^{\lambda_k}_{\mu}$, $\lambda_k=0,\pm 1$, 
given in Appendix~\ref{Representation of momentum and spin vectors}.
Then we obtain for the decay matrices 
\begin{eqnarray} \label{rhoD1}
\rho_{D_1}(\tilde\chi^+_i)_{\lambda_i' \lambda_i}^{\lambda_k\lambda'_k} &=& 
(\delta_{\lambda_i' \lambda_i} D_1^{\mu\nu} 
+ \sigma^a_{\lambda_i'\lambda_i}  \Sigma^{a~\mu\nu}_{D_1})
		  \varepsilon_{\mu}^{\lambda_k\ast}\varepsilon_{\nu}^{\lambda'_k}
\end{eqnarray}
and
\begin{eqnarray}\label{rhoD2}
		  \rho_{D_2}(W^+)_{\lambda'_k\lambda_k}&=& D_2^{\mu\nu}
\varepsilon_{\mu}^{\lambda_k}\varepsilon_{\nu}^{\lambda'_k\ast},
\end{eqnarray}
with:
\begin{eqnarray}
	D_1^{\mu\nu} &=& g^2(|O^{R}_{ni}|^2+|O^{L}_{ni}|^2)
	[2 p^{\mu}_{\chi^+_i} p^{\nu}_{\chi^+_i}
	 -(p^{\mu}_{\chi^+_i} p^{\nu}_W + p^{\nu}_{\chi^+_i} p^{\mu}_W)
	 -{\textstyle\frac{1}{2}}(m_{\chi^+_i}^2+m_{\chi^0_n}^2-m_W^2)
	 g^{\mu\nu} ] \nonumber\\
 && +2g^2 Re(O^{R\ast}_{ni}O^{L}_{ni})
	m_{\chi^+_i}m_{\chi^0_n}g^{\mu\nu} 
	\,^{\;\,+}_{(-)}  
	ig^2(|O^{R}_{ni}|^2-|O^{L}_{ni}|^2)\epsilon^{\mu\alpha\nu\beta}
          p_{\chi^+_i\alpha}p_{W\beta} , \\
 \Sigma_{D_1}^{a~\mu\nu}&=&
			 \,^{\;\,+}_{(-)}
			 g^2(|O^{R}_{ni}|^2-|O^{L}_{ni}|^2)m_{\chi^+_i}
			 [s^{a,\mu}_{\chi^+_i}(p^{\nu}_{\chi^+_i} -p^{\nu}_{W})
			+s^{a,\nu}_{\chi^+_i}(p^{\mu}_{\chi^+_i}-p^{\mu}_{W})
			+(s^{a}_{\chi^+_i}\cdot p_{W})g^{\mu\nu}]\nonumber\\
	&&-ig^2(|O^{R}_{ni}|^2+|O^{L}_{ni}|^2)m_{\chi^+_i}
		\epsilon^{\mu\alpha\nu\beta}s^a_{\chi^+_i\alpha}
		(p_{\chi^+_i\beta}-p_{W\beta})\nonumber\\
	&&+2ig^2 Re(O^{R\ast}_{ni}O^{L}_{ni})m_{\chi^0_n}
		\epsilon^{\mu\alpha\nu\beta}s^a_{\chi^+_i\alpha}
		p_{\chi^+_i\beta}\nonumber\\
&&-2ig^2 Im(O^{R\ast}_{ni}O^{L}_{ni})m_{\chi^0_n}
			(s^{a,\mu}_{\chi^+_i}p^{\nu}_{\chi^+_i}-
			s^{a,\nu}_{\chi^+_i}p^{\mu}_{\chi^+_i}	)
	; \quad (\epsilon_{0123}=1),
\end{eqnarray}
and
\begin{eqnarray}
	D_2^{\mu\nu} &=&g^2(-2p^{\mu}_{\bar f}p^{\nu}_{\bar f} 
		+p^{\mu}_Wp^{\nu}_{\bar f} +p^{\mu}_{\bar f}p^{\nu}_W
		-{\textstyle\frac{1}{2}}m_W^2 g^{\mu\nu})
\,^{\;\,-}_{(+)}ig^2\epsilon^{\mu\alpha\nu\beta}p_{W\alpha}p_{\bar f\beta},
\end{eqnarray}
where here, and in the following, 
the signs in parenthesis hold for the charge conjugated processes, 
here $\tilde\chi^-_i\to W^-\tilde\chi^0_n$ and
$W^- \to \bar f^{'} f$, respectively.
In (\ref{rhoD1}) and (\ref{rhoD2}) we use the 
expansion \cite{choiBM}:
\begin{eqnarray}\label{expansion}
\varepsilon_{\mu}^{\lambda_k}\varepsilon_{\nu}^{\lambda'_k\ast}&=&
{\textstyle \frac{1}{3}}\delta^{\lambda_k'\lambda_k}I_{\mu\nu}
-\frac{i}{2m_W}\epsilon_{\mu\nu\rho\sigma}
p_W^{\rho}t_W^{c\sigma}(J^c)^{\lambda_k'\lambda_k}
-{\textstyle \frac{1}{2}}t_{W\mu}^c t_{W\nu}^d (J^{cd})^{\lambda_k'\lambda_k},
\end{eqnarray}
summed over $c,d$, and $\epsilon_{0123}=1$. $J^c$, $c=1,2,3$, are the 
$3\times3$ spin-1 matrices with
$	[J^c,J^d]=i\epsilon_{cde}J^e.$
The matrices
\begin{eqnarray}
	J^{cd}&=&J^cJ^d+J^dJ^c-{\textstyle \frac{4}{3}}\delta^{cd},
\end{eqnarray}
with $J^{11}+J^{22}+J^{33}=0$,
are the components of a symmetric, traceless tensor.
An explicit form of $J^c$  and $J^{cd}$ is given in Appendix~\ref{matrices}.
The completeness relation of the polarization vectors
\begin{eqnarray}\label{completeness}
\sum_{\lambda_k} \varepsilon^{\lambda_k\ast}_{\mu}
\varepsilon^{\lambda_k}_{\nu}&=& -g_{\mu\nu}+\frac{p_{W \mu}p_{W \nu}}{m_W^2}
\end{eqnarray}
is guaranteed by
\begin{eqnarray}
I_{\mu\nu}&=&-g_{\mu\nu}+\frac{p_{W \mu}p_{W \nu}}{m_W^2}.
\end{eqnarray}
The second term of (\ref{expansion}) describes the vector
polarization and the third term describes the tensor polarization
of the $W$ boson.
The decay matrices can be expanded in terms of the spin matrices
$J^c$ and $J^{cd}$. The first term  of the decay
matrix $\rho_{D_1}$ (\ref{rhoD1}), 
which is independent of the chargino polarization, then is
\begin{eqnarray}
D_1^{\mu\nu} 
\varepsilon_{\mu}^{\lambda_k\ast}\varepsilon_{\nu}^{\lambda'_k}&=&
  D_1 \delta^{\lambda_k\lambda_k'} 
  + \,^cD_1(J^c)^{\lambda_k\lambda_k'}
  + \,^{cd}D_1(J^{cd})^{\lambda_k\lambda_k'},
\end{eqnarray}
summed over $c,d$, with
\begin{eqnarray}\label{D1}
	D_1&=&{\textstyle \frac{1}{6}}g^2(|O^{R}_{ni}|^2+|O^{L}_{ni}|^2)
	\Big[m_{\chi^+_i}^2+m_{\chi^0_n}^2-2 m_W^2
		+\frac{(m_{\chi^+_i}^2-m_{\chi^0_n}^2)^2}{m_W^2}\Big]
	\nonumber\\
&&-2g^2Re( O^{R\ast}_{ni}O^{L}_{ni})m_{\chi^+_i}m_{\chi^0_n},\\
^{c}D_1&=&\,^{\;\,+}_{(-)}g^2(|O^{R}_{ni}|^2-|O^{L}_{ni}|^2)m_W(t^c_W\cdot
         	p_{\chi^+_i}),\label{cD1}\\
^{cd}D_1&=&-g^2 (|O^{R}_{ni}|^2+|O^{L}_{ni}|^2)
			\left[(t^c_W\cdot p_{\chi^+_i})(t^d_W\cdot p_{\chi^+_i})+
	{\textstyle \frac{1}{4}}(m_{\chi^+_i}^2+m_{\chi^0_n}^2-m_W^2)
		\delta^{cd}\right]\nonumber \\
&& + g^2Re( O^{R\ast}_{ni}O^{L}_{ni})m_{\chi^+_i}m_{\chi^0_n}
		\delta^{cd}.\label{cdD1}
\end{eqnarray}
As a consequence of the completeness relation
(\ref{completeness}), the diagonal coefficients are linearly
dependent
\begin{eqnarray}
	^{11}D_1+\,^{22}D_1+\,^{33}D_1&=&-{\textstyle \frac{3}{2}}D_1.
\end{eqnarray}
For large chargino momentum ${\mathbf p}_{\chi^+_i}$, the $W$ boson will
mainly be emitted in the direction of 
${\mathbf p}_{\chi^+_i}$, i.e. 
$\hat {\mathbf p}_{\chi^+_i}\approx\hat {\mathbf p}_{W}$, 
with $\hat {\mathbf p}={\mathbf p}/|{\mathbf p}|$.
Therefore, for high energies we have 
$(t^{1,2}_W\cdot p_{\chi^+_i}) \approx 0$ in (\ref{cdD1}),
resulting in $^{11}D_1\approx \,^{22}D_1$.

For the second term of $\rho_{D_1}$ (\ref{rhoD1}), 
which depends on the polarization of the decaying chargino,
we obtain
\begin{eqnarray}
\Sigma^{a~\mu\nu}_{D_1}
\varepsilon_{\mu}^{\lambda_k\ast}\varepsilon_{\nu}^{\lambda'_k}&=&
  \Sigma^{a}_{D_1}\delta^{\lambda_k\lambda_k'}
  +\,^c\Sigma^{a}_{D_1}(J^c)^{\lambda_k\lambda_k'}
  + \,^{cd}\Sigma^{a}_{D_1}(J^{cd})^{\lambda_k\lambda_k'},
\end{eqnarray}
summed over $c$, $d$, with
\begin{eqnarray}
\Sigma^{a}_{D_1}&=&\,^{\;\,+}_{(-)}
	{\textstyle \frac{2}{3}}g^2 (|O^{R}_{ni}|^2-|O^{L}_{ni}|^2)
		m_{\chi^+_i}(s^a_{\chi^+_i}\cdot p_W)[
			\frac{m_{\chi^+_i}^2-m_{\chi^0_n}^2}{2 m_W^2}-1],
		\label{SigmaaD1}\\
^c\Sigma^{a}_{D_1}&=&\frac{g^2}{m_W}(|O^{R}_{ni}|^2+|O^{L}_{ni}|^2)
			m_{\chi^+_i}\left[
			(t^c_W\cdot p_{\chi^+_i})(s^a_{\chi^+_i}\cdot p_W)
			+{\textstyle \frac{1}{2}}(t^c_W\cdot s^a_{\chi^+_i})
			(m^2_{\chi^0_n}-m^2_{\chi^+_i}+m_W^2)\right]
		\nonumber\\
		&&-\frac{2g^2}{m_W}Re( O^{R\ast}_{ni}O^{L}_{ni})
			m_{\chi^0_n}\left[
			(t^c_W\cdot p_{\chi^+_i})(s^a_{\chi^+_i}\cdot p_W)
			+{\textstyle \frac{1}{2}}(t^c_W\cdot s^a_{\chi^+_i})
			(m^2_{\chi^0_n}-m^2_{\chi^+_i}-m_W^2)\right]
		\nonumber\\
		&&+\frac{2g^2}{m_W}Im( O^{R\ast}_{ni}O^{L}_{ni})
		m_{\chi^0_n}\epsilon_{\mu\nu\rho\sigma}
		s^{a\mu}_{\chi^+_i}p^{\nu}_{\chi^+_i}
		p^{\rho}_W t^{c\sigma}_W, \label{csigmaaD1}\\
^{cd}\Sigma^{a}_{D_1}&=&\,^{\;\,+}_{(-)}{\textstyle \frac{1}{2}}g^2
		(|O^{R}_{ni}|^2-|O^{L}_{ni}|^2)m_{\chi^+_i} \times \nonumber\\
&&\left[(s^a_{\chi^+_i}\cdot p_W)\delta^{cd}
			-(t^c_W\cdot p_{\chi^+_i})(t^d_W\cdot s^a_{\chi^+_i})
			-(t^d_W\cdot p_{\chi^+_i})(t^c_W\cdot s^a_{\chi^+_i})
			\right].
\end{eqnarray}
A similar expansion for 
the $W$ decay matrix (\ref{rhoD2}), results in
\begin{eqnarray}
	 \rho_{D_2}(W^+)_{\lambda'_k\lambda_k}&=&   
D_2 \delta^{\lambda_k'\lambda_k} 
  + \,^cD_2(J^c)^{\lambda_k'\lambda_k}
  + \,^{cd}D_2(J^{cd})^{\lambda_k'\lambda_k},
\end{eqnarray}
where we sum over $c$, $d$, with 
\begin{eqnarray}
D_2&=&{\textstyle\frac{1}{3}}g^2 m_W^2,\label{D2}\\
^{c}D_2&=&\,^{\;\,-}_{(+)}g^2 m_W(t^c_W\cdot p_{\bar f}),\label{cD2}\\
^{cd}D_2&=&g^2 \left[(t^c_W\cdot p_{\bar f})(t^d_W\cdot p_{\bar f})-
	{\textstyle \frac{1}{4}}m_W^2\delta^{cd}\right],\label{cdD2}
\end{eqnarray}
The diagonal coefficients are linearly dependent
\begin{eqnarray}
	^{11}D_2+\,^{22}D_2+\,^{33}D_2&=&-{\textstyle \frac{3}{2}}D_2.
\end{eqnarray}

Inserting the density matrices (\ref{rhoP}) and (\ref{rhoD1}) 
into (\ref{Wdensitymatrix}) leads to:
\begin{eqnarray}\label{Wdensitymatrixunnorm}
\rho_{P}(W^+)^{\lambda_k\lambda'_k}&=&
4~|\Delta(\tilde\chi^+_i)|^2~[
	(P  D_1 + \Sigma^{a}_{P}\Sigma^{a}_{D_1}) ~\delta^{\lambda_k\lambda_k'}
	+(P \;^{c}D_1+\Sigma_P^a \,^{c}\Sigma^{a}_{D_1})
	~(J^{c})^{\lambda_k\lambda_k'}\nonumber\\
&&+(P  \;^{cd}D_1 +\Sigma^{a}_{P} \;^{cd}\Sigma^{a}_{D_1})
	~(J^{cd})^{\lambda_k\lambda_k'}],
\end{eqnarray}
summed over $a,c,d$. Inserting then (\ref{Wdensitymatrixunnorm}) 
and (\ref{rhoD2}) into (\ref{amplitude}) leads to
the decomposition of the amplitude 
squared in its scalar (first term), vector (second term)
and tensor part (third term):
\begin{eqnarray} \label{amplitude2}
|T|^2 &=& 4~|\Delta(\tilde\chi^+_i)|^2~ |\Delta(W^+)|^2
	\{3( P  D_1 + \Sigma^{a}_{P}\Sigma^{a}_{D_1} )D_2 +  
		2(P \;^{c}D_1+\Sigma_P^a \,^c\Sigma_{D_1}^a) \,^cD_2 \nonumber\\
&& \quad \quad  
		+4[(P \;^{cd}D_1  +\Sigma^{a}_{P} \;^{cd}\Sigma^{a}_{D_1} )\,^{cd} D_2-
		{\textstyle \frac{1}{3}}
		(P\;^{cc}D_1 +\Sigma^{a}_{P}\;^{cc}\Sigma^{a}_{D_1})\,^{dd} D_2]\},
\end{eqnarray}
summed over $a,c,d$.

\subsection{Density matrix of the $W$ boson
     \label{Density matrix}}

The mean polarization of the $W$ bosons in the laboratory system 
is given by the $3\times3$ density matrix $<\rho(W^+)>$,
obtained by integrating (\ref{Wdensitymatrixunnorm}) 
over the Lorentz invariant phase space element 
$d{\rm Lips}(s,p_{\chi^-_j },p_{\chi^0_n},p_{W})$
see (\ref{Lips}), and normalizing by the trace:
\begin{equation}\label{Wdensitymatrixnorm}
<\rho(W^+)^{\lambda_k\lambda'_k}>=
\frac{\int \rho_{P}(W^+)^{\lambda_k\lambda'_k}~d{\rm Lips}}
		{\int {\rm Tr} \{\rho_{P}(W^+)^{\lambda_k\lambda'_k}\}~d{\rm Lips}}
={\textstyle \frac{1}{3}}\delta^{\lambda_k\lambda_k'}
+V_c ~(J^{c})^{\lambda_k\lambda_k'}
+T_{cd}  ~(J^{cd})^{\lambda_k\lambda_k'},\label{defcoef}
\end{equation}
summed over $c$, $d$. The vector and tensor coefficients
$V_c$ and $T_{cd}$ are given by:
\begin{eqnarray}\label{Vs}
	V_c&=&\frac{\int |\Delta(\tilde\chi^+_i)|^2
		~(P \, ^cD_1 + \Sigma_P^a \,^{c}\Sigma^{a}_{D_1}) 
		~d{\rm Lips}}
{3 \int |\Delta(\tilde\chi^+_i)|^2 ~P  D_1~d{\rm Lips}},\\
T_{cd}&=&T_{dc}\;=\;
\frac{\int |\Delta(\tilde\chi^+_i)|^2  
	~(P  \;^{cd}D_1+\Sigma^{a}_{P} \;^{cd}\Sigma^{a}_{D_1} )
		~d{\rm Lips}}
{3 \int |\Delta(\tilde\chi^+_i)|^2 ~P  D_1~d{\rm Lips}},\label{Ts}
\end{eqnarray}
with sum over a. 
The density matrix in the circular polarization basis 
(\ref{circularbasis}) is given  by 
\begin{eqnarray} \label{density1}
	<\rho(W^+)^{--}> &= &
	{\textstyle \frac{1}{2}}-V_3+T_{33}, \\
	<\rho(W^+)^{00}> &=&-2T_{33},\\
	<\rho(W^+)^{-0}> &= &
	{\textstyle \frac{1}{\sqrt{2}}}(V_1+iV_2)-\sqrt{2}\,(T_{13}+iT_{23}),\\
		<\rho(W^+)^{-+}> &= & T_{11}-T_{23}+2iT_{12},\\
	<\rho(W^+)^{0+}> &= & {\textstyle \frac{1}{\sqrt{2}}}(V_1+iV_2)
	+\sqrt{2}\,(T_{13}+iT_{23}),\label{density5}
\end{eqnarray}
where we have used $T_{11}+T_{22}+T_{33}=-\frac{1}{2}$.

\section{T odd asymmetries
	\label{T odd asymmetries}}

From (\ref{Wdensitymatrixunnorm}) we obtain for asymmetry  
${\mathcal A}_{I}^{\rm T}$  (\ref{AT1}):
\begin{equation} 
	 {\mathcal A}_{I}^{\rm T} 
	 = \frac{\int {\rm Sign}[{\mathcal T}_{I}]
		 {\rm Tr} \{\rho_{P}(W^+)^{\lambda_k\lambda'_k}\} d{\rm Lips}}
	 {\int {\rm Tr} \{\rho_{P}(W^+)^{\lambda_k\lambda'_k}\} d{\rm Lips}}
	 =
 \frac{\int |\Delta(\tilde\chi^+_i)|^2 ~
	{\rm Sign}[{\mathcal T}_{I}]
		\Sigma_P^2 \, \Sigma_{D_1}^2  d{\rm Lips}}
	{\int  |\Delta(\tilde\chi^+_i)|^2~ P D_1 d{\rm Lips}},
 \label{asymI}
\end{equation}
with
$d{\rm Lips}(s,p_{\chi^-_j },p_{\chi^0_n},p_{W})
$$~d s_{\chi^+_i} \,\sum_{\pm}
$$d{\rm Lips}(s_{\chi^+_i},p_{\chi^0_n},p_{W}^{\pm})$,
given in (\ref{Lips}).
In the numerator of (\ref{asymI}), only the spin correlations
$\Sigma_P^2 \, \Sigma_{D_1}^2$ perpendicular to the production plane
remain, since only this term contains the  the triple product
${\mathcal T}_{I} ={\mathbf p}_{e^-}\cdot({\mathbf p}_{\chi^+_i} 
\times {\mathbf p}_{W})$. In the denominator only the term $P D_1$ remains,
and all spin correlations vanish due to the integration over the
complete phase space. Note that 
${\mathcal A}_{I}^{\rm T}\propto
\Sigma_{D_1}^2\propto(|O^{R}_{ni}|^2-|O^{L}_{ni}|^2)$,
see  (\ref{SigmaaD1}), and thus ${\mathcal A}_{I}^{\rm T}$ may be reduced
for $|O^{R}_{ni}| \approx |O^{L}_{ni}|$.
Moreover ${\mathcal A}_{I}^{\rm T}$ will be small for
$m_{\chi^+_i}^2-m_{\chi^0_n}^2 \approx 2 m_W^2$.

For the asymmetry  ${\mathcal A}_{II}^{\rm T}$ (\ref{AT2}),
we obtain from (\ref{amplitude2}): 
\begin{equation} 
	 {\mathcal A}_{II}^{\rm T} 
	 = \frac{\int {\rm Sign}[{\mathcal T}_{II}]
		 |T|^2 d{\rm Lips}}
           {\int |T|^2 d{\rm Lips}}  
 = \frac{\int |\Delta(\tilde\chi^+_i)|^2 |\Delta(W^+)|^2~
	 {\rm Sign}[{\mathcal T}_{II}]
	2\Sigma_P^a \, ^c\Sigma_{D_1}^a \,^cD_2 d{\rm Lips}}
	{\int |\Delta(\tilde\chi^+_i)|^2 |\Delta(W^+)|^2~
		3 P D_1D_2 d{\rm Lips}}, \label{asymII}
\end{equation}
summed over $a$ and $c$, with 
$d{\rm Lips}= d{\rm Lips}(s,p_{\chi^-_j },
p_{\chi^0_n},p_{f^{'}},p_{\bar f})$, defined in  (\ref{Lips}).
In the numerator only the vector part of $|T|^2$ 
remains because only the vector part contains the triple product
${\mathcal T}_{II}={\mathbf p}_{e^-}\cdot({\mathbf p}_{c} 
			 \times {\mathbf p}_{\bar s}) $.
In the denominator the vector and tensor parts of $|T|^2$ vanish
due to phase space integration. Owing to the correlations between
the $\tilde\chi^+_i$ and the $W$ boson polarization, 
$\Sigma_P^a \, ^c\Sigma_{D_1}^a$, there are 
 contributions to the asymmetry ${\mathcal A}_{II}^{\rm T}$ 
from the chargino production process (\ref{production}),
and/or from the chargino decay process (\ref{decay_1}).
The contribution from the production is given by the term with $a=2$ 
in (\ref{asymII}) and it is proportional to 
the transverse polarization of the chargino perpendicular to the 
production plane, $\Sigma^{2}_P$ (\ref{rhoP}). 
For $e^+e^- \to\tilde\chi^+_i \tilde\chi^+_i$ we have $\Sigma^{2}_P=0$.
The contributions  from the decay, which are the terms with $a=1,3$
in (\ref{asymII}), are proportional to  
\begin{eqnarray}\label{adependence}
	^c\Sigma_{D_1}^a \,^cD_2&\supset&
	-2g^4 m_{\chi_n^0}
	Im( O^{R\ast}_{ni}O^{L}_{ni})(t^c_W\cdot p_{\bar f})
	\epsilon_{\mu\nu\rho\sigma}s^{a,\mu}_{\chi^+_i}p_{\chi^+_i}^{\nu}
	p_W^{\rho}t^{c\sigma}_W,
\end{eqnarray}
which contains the $\epsilon$-tensor,
see the last term of (\ref{csigmaaD1}). Thus ${\mathcal A}_{II}^{\rm T}$ 
can be enhanced (reduced) if the contributions 
from production and decay  have the same (opposite) sign.
Note that the contributions from the decay would vanish 
for a two-body decay of the chargino into a scalar particle
instead of a $W$ boson.

The relative statistical error 
of ${\mathcal A}_{I}^{\rm T}$ is $\delta {\mathcal A}_{I}^{\rm T} = 
\Delta {\mathcal A}_{I}^{\rm T}/|{\mathcal A}_{I}^{\rm T}| = 
1/(|{\mathcal A}_{I}^{\rm T}| \sqrt{N})$, 
where $N={\mathcal L} \cdot\sigma$ is the number of events for the integrated 
luminosity ${\mathcal L}$ and the cross section 
$\sigma=\sigma_P(e^+e^-\to\tilde\chi^+_i\tilde\chi^-_j) 
\times{\rm BR}(\tilde\chi^+_i \to W^+\tilde\chi^0_n)$.
For the CP asymmetry ${\mathcal A}_{I}$ (\ref{ACP}), we have
$\Delta {\mathcal A}_{I}=\Delta {\mathcal A}_{I}^{\rm T}/\sqrt{2} $.
The statistical significance, with which the the asymmetry can be
measured, is then given by
$S_{I} = |{\mathcal A}_{I}| \sqrt{2{\mathcal L}\cdot\sigma}$.
A similar result is obtained for 
the asymmetry ${\mathcal A}_{II}$ with 
$S_{II} = |{\mathcal A}_{II}| 
\sqrt{2{\mathcal L}\cdot\sigma}$ and the cross section
$\sigma=\sigma_P(e^+e^-\to\tilde\chi^+_i\tilde\chi^-_j) 
\times{\rm BR}(\tilde\chi^+_i \to W^+\tilde\chi^0_n)\times
{\rm BR}(W^+\to c\bar s)$.
Note that in order to measure ${\mathcal A}_{I}$ 
the momentum of $\tilde\chi^+_i$, i.e. the
production plane, has to be kinematically reconstructed.
This could be accomplished by measuring the decay of the 
other chargino $\tilde\chi^-_j$, 
if the masses of the charginos and the masses of their decay 
products are known.
For the measurement of ${\mathcal A}_{II}$, the flavors of the 
quarks $c$ and $\bar s$ have to be  distinguished, which 
will be possible by flavor tagging of the 
$c$-quark \cite{flavortaggingatLC,flavortagginginWdecays}. 
In principle, for the decay  $W \to u ~\bar d$ also an asymmetry similar 
to ${\mathcal A}_{II}$ can be considered, if it is possible to
distinguish between the $u$ and $\bar d$ jet, for instance,
by measuring the average charge.
Also it is clear that detailed
Monte Carlo studies taking into account background and detector
simulations are necessary to predict the expected accuracies.
However, this is beyond the scope of the present work.

\section{Numerical results
	\label{Numerical results}}

We study the dependence of  
${\mathcal A}_{I}$,  ${\mathcal A}_{II}$ (\ref{ACP}), and 
the density matrix $<\rho(W^+)>$ (\ref{Wdensitymatrixnorm}),
on the MSSM parameters 
$\mu = |\mu| \, e^{ i\,\varphi_{\mu}}$, 
$M_1 = |M_1| \, e^{ i\,\varphi_{M_1}}$,
$\tan \beta$ and 
	the universal scalar mass parameter
$m_0$. 
We will allow $\varphi_{M_1}\in[\pi,-\pi ]$, however take into account 
$|\varphi_{\mu}|\lsim 0.1 \pi$
in some of the plots,
as suggested  from the EDM analyses \cite{nath,edms}. 
In order to show the full phase dependence 
of the asymmetries, however, 
we will relax the EDM restrictions in some 
	of the examples studied.

The feasibility of measuring the asymmetries depends also on the cross 
sections $\sigma=\sigma_P(e^+e^-\to\tilde\chi^+_i\tilde\chi^-_j ) \times
{\rm BR}( \tilde\chi^+_i \to W^+\tilde\chi^0_1)\times
{\rm BR}(W^+\to c \bar s)$, 
which we will discuss in our scenarios. 
We choose  a center  of mass energy of   $\sqrt{s} = 800$ GeV
and longitudinally polarized beams with $(P_{e^-},P_{e^+})=(-0.8,+0.6)$.
This choice enhances sneutrino exchange in the chargino 
production process, which results in larger cross sections 
and asymmetries. For the calculation of the  branching ratios 
${\rm BR}( \tilde\chi^+_i \to W^+\tilde\chi^0_1)$
and widths $\Gamma_{\chi_i^+}$, we include the two-body decays:
\begin{eqnarray}
	\tilde\chi^+_i &\to& 
	W^+\tilde\chi^0_n,~
	\tilde e_{L}^+\nu_{e},~
	\tilde\mu_{L}^+\nu_{\mu},~
	\tilde\tau_{1,2}^+\nu_{\tau},~
	e^+\tilde\nu_{e},~
	\mu^+\tilde\nu_{\mu},~
	\tau^+\tilde\nu_{\tau},
\end{eqnarray}
and neglect three-body decays. 
For the $W$ boson decay
we take  the experimental value
${\rm BR}(W^+\to c \bar s)=0.31$  \cite{PDG}.
In order to reduce the number of parameters, we assume the 
relation $|M_1|=5/3 \, M_2\tan^2\theta_W $ and  
use the renormalization group equations \cite{hall} for the 
slepton and sneutrino masses,
$m_{\tilde\ell_L  }^2 = m_0^2 +0.79 M_2^2
+m_Z^2\cos 2 \beta(-1/2+ \sin^2 \theta_W)$ and 
$m_{\tilde\nu_{\ell}  }^2 = m_0^2 +0.79 M_2^2
+m_Z^2/2\cos 2 \beta$.
In the stau sector \cite{thomas}, we fix the trilinear scalar coupling
parameter to $A_{\tau}=250$ GeV.

\subsection{Production of $\tilde\chi^+_1 \, \tilde\chi^-_1$ }

\begin{figure}[t]
\setlength{\unitlength}{1cm}
\begin{picture}(10,8)(-0.5,0)
	\put(0,0){\includegraphics{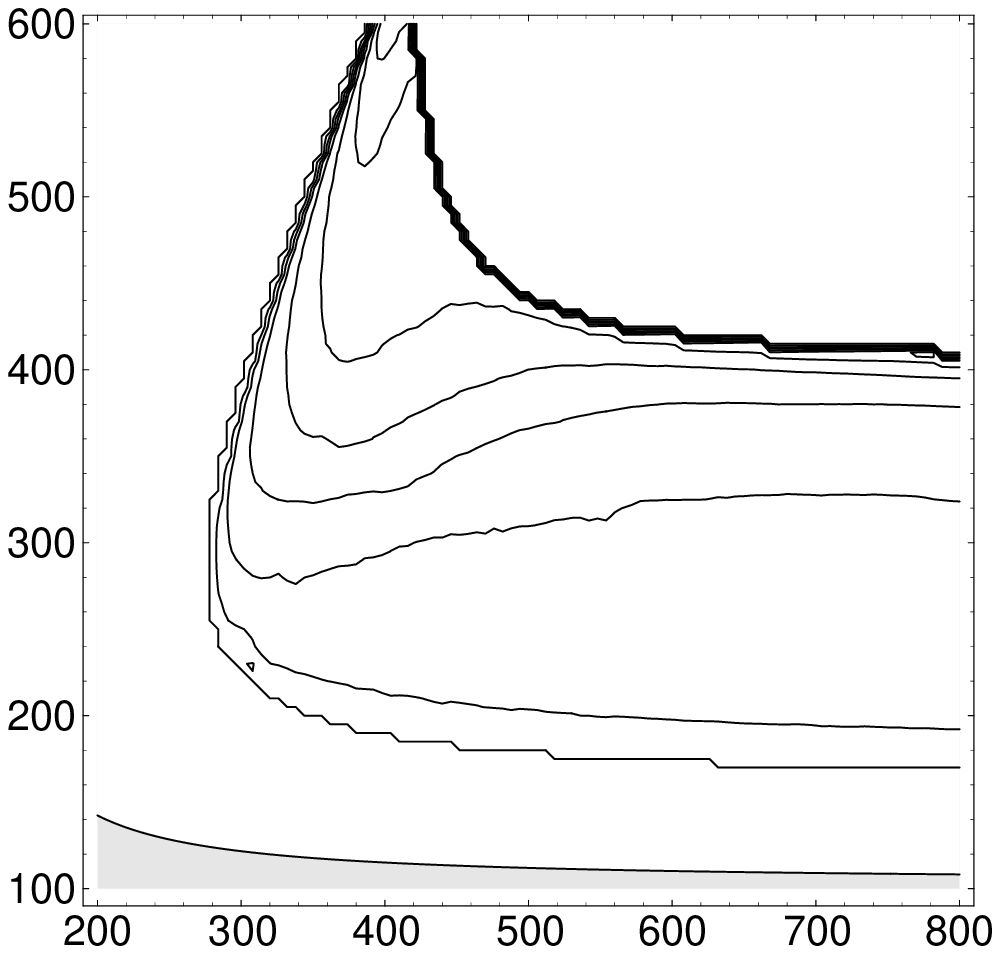}}
		\put(3.,7.4){\fbox{${\mathcal A}_{II}$ in \%}}
		\put(5.5,-0.3){$|\mu|\,[{\rm GeV}]$}
	\put(0,7.4){$M_2\,[{\rm GeV}]$ }
	\put(2.7,6.35){\scriptsize 7}
	\put(2.6,5.55){\footnotesize 6}
	\put(2.8,4.35){\footnotesize 5}
	\put(3.,3.85){\footnotesize 4}
	\put(3.2,3.45){\footnotesize 3}
	\put(3.5,2.9){\footnotesize 2}
	\put(4.,1.65){\footnotesize 1}
	\put(5.5,6){\begin{picture}(1,1)(0,0)
			\CArc(0,0)(7,0,380)
			\Text(0,0)[c]{{\footnotesize A}}
	\end{picture}}
			\put(1.3,5.0){\begin{picture}(1,1)(0,0)
			\CArc(0,0)(7,0,380)
			\Text(0,0)[c]{{\footnotesize B}}
		\end{picture}}
\put(0.5,-.3){Fig.~\ref{plot2}a}
\put(8,0){\includegraphics{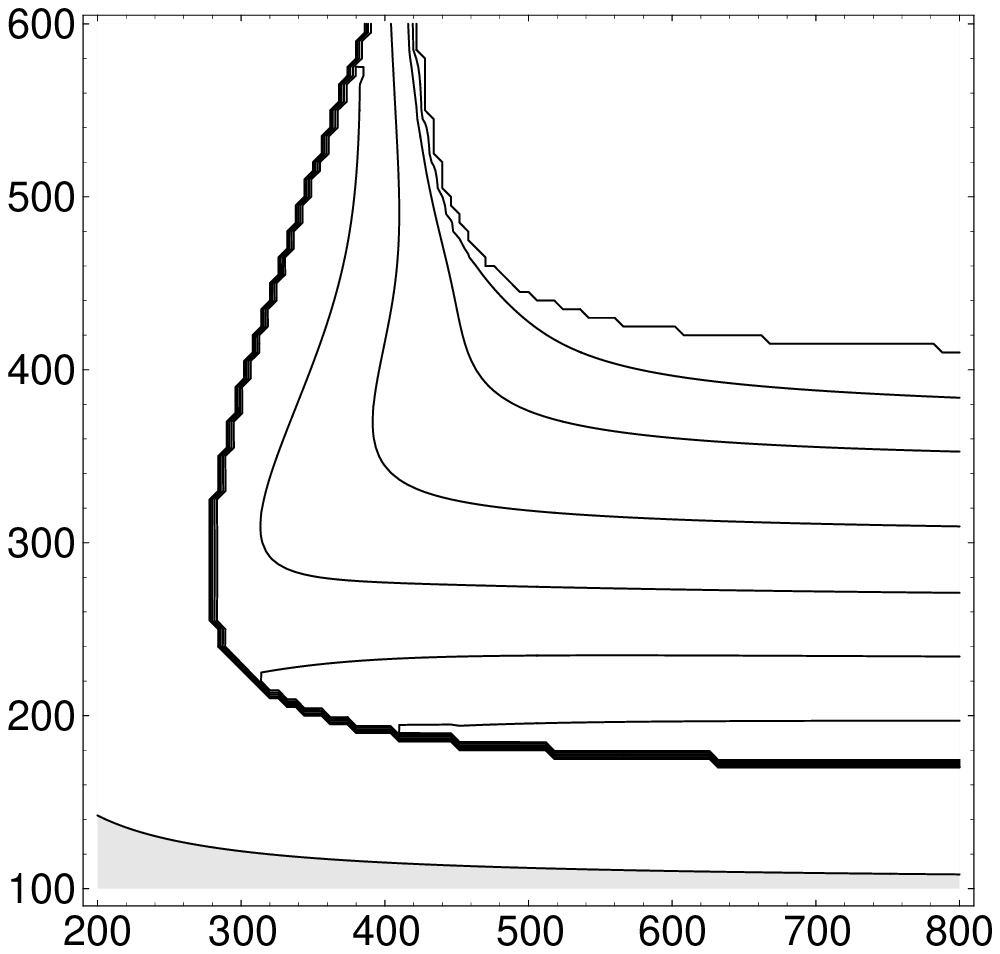}}
	\put(11.5,7.4){\fbox{$\sigma$ in fb}}
	\put(13.5,-.3){$|\mu|\,[{\rm GeV}]$}
	\put(8,7.4){$M_2\,[{\rm GeV}]$ }
	\put(14.0,4.25){\footnotesize 10}
	\put(13.6,3.9){\footnotesize 25}
	\put(12.9,3.4){\footnotesize 50}
	\put(12.5,2.9){\footnotesize 75}
	\put(12.0,2.4){\footnotesize 100}
	\put(11.5,1.9){\footnotesize 125}
	  	\put(13.5,6){\begin{picture}(1,1)(0,0)
			\CArc(0,0)(7,0,380)
			\Text(0,0)[c]{{\footnotesize A}}
	\end{picture}}
			\put(9.3,5.0){\begin{picture}(1,1)(0,0)
			\CArc(0,0)(7,0,380)
			\Text(0,0)[c]{{\footnotesize B}}
		\end{picture}}
	\put(8.5,-.3){Fig.~\ref{plot2}b}
\end{picture}
\vspace*{.5cm}
\caption{
	Contour lines of
	the asymmetry ${\mathcal A}_{II}$ (\ref{plot2}a)
	and $\sigma=\sigma_P(e^+e^-\to\tilde\chi^+_1\tilde\chi^-_1) 
	\times {\rm BR}( \tilde\chi^+_1 \to W^+\tilde\chi^0_1)
	\times {\rm BR}(W^+ \to c \bar s)$ (\ref{plot2}b),
	in the $|\mu|$--$M_2$ plane for 
	$(\varphi_{M_1},\varphi_{\mu})=(0.5\pi,0)$,
	$\tan \beta=5$, $m_0=300$ GeV,
	$\sqrt{s}=800$ GeV and $(P_{e^-},P_{e^+})=(-0.8,0.6)$.
	The area A is kinematically forbidden by
	$m_{\chi^+_1}+m_{\chi^-_1}>\sqrt{s}$,
	the area B by
	$m_{W}+m_{\chi^0_1}> m_{\chi^+_1}$.
	The gray  area is excluded by $m_{\chi_1^{\pm}}<104 $ GeV.
	\label{plot2}}
\end{figure}

For the production $e^+e^-\to\tilde\chi^+_i\tilde\chi^-_i $
of a pair of charginos the polarization perpendicular to the production
plane vanishes, and thus ${\mathcal A}_{I}=0$.
However, ${\mathcal A}_{II}$ need not to be zero and is 
sensitive to $\varphi_{\mu}$ and $\varphi_{M_1}$, because this
asymmetry has contributions from the chargino decay process.
For $(\varphi_{M_1},\varphi_{\mu})=(0.5\pi,0)$ we show 
in Fig.~\ref{plot2}a the $|\mu|$--$M_2$ dependence of 
${\mathcal A}_{II}$, which can reach values of 5\%-7\% 
for $M_2 \gsim 400$~GeV. We also studied the $\varphi_{\mu}$ dependence of 
${\mathcal A}_{II}$ in the $|\mu|$--$M_2$ plane. For
$\varphi_{M_1}=0$, $\varphi_{\mu}=0.1\pi(0.5\pi)$ 
and the other parameters as given in the caption of Fig.~\ref{plot2},
we find $|{\mathcal A}_{II}|<2\%(7\%)$. 

In Fig.~\ref{plot2}b we show the contour lines of the
cross section 
$\sigma=\sigma_P(e^+e^-\to\tilde\chi^+_1\tilde\chi^-_1 ) \times
{\rm BR}( \tilde\chi^+_1 \to W^+\tilde\chi^0_1)\times
{\rm BR}(W^+ \to c \bar s)$ in the $|\mu|$--$M_2$ plane for 
$(\varphi_{M_1},\varphi_{\mu})=(0.5\pi,0)$.
The production cross section 
$\sigma_P(e^+e^-\to\tilde\chi^+_1\tilde\chi^-_1 )$
reaches more than $400$~fb. For our choice
of $m_0=300$ GeV, $ \tilde\chi^+_1 \to W^+\tilde\chi^0_1$
is the only allowed two-body decay  channel.
%

\begin{figure}[t]
\setlength{\unitlength}{1cm}
\begin{picture}(10,8)(-0.5,0)
   \put(0,0){\includegraphics{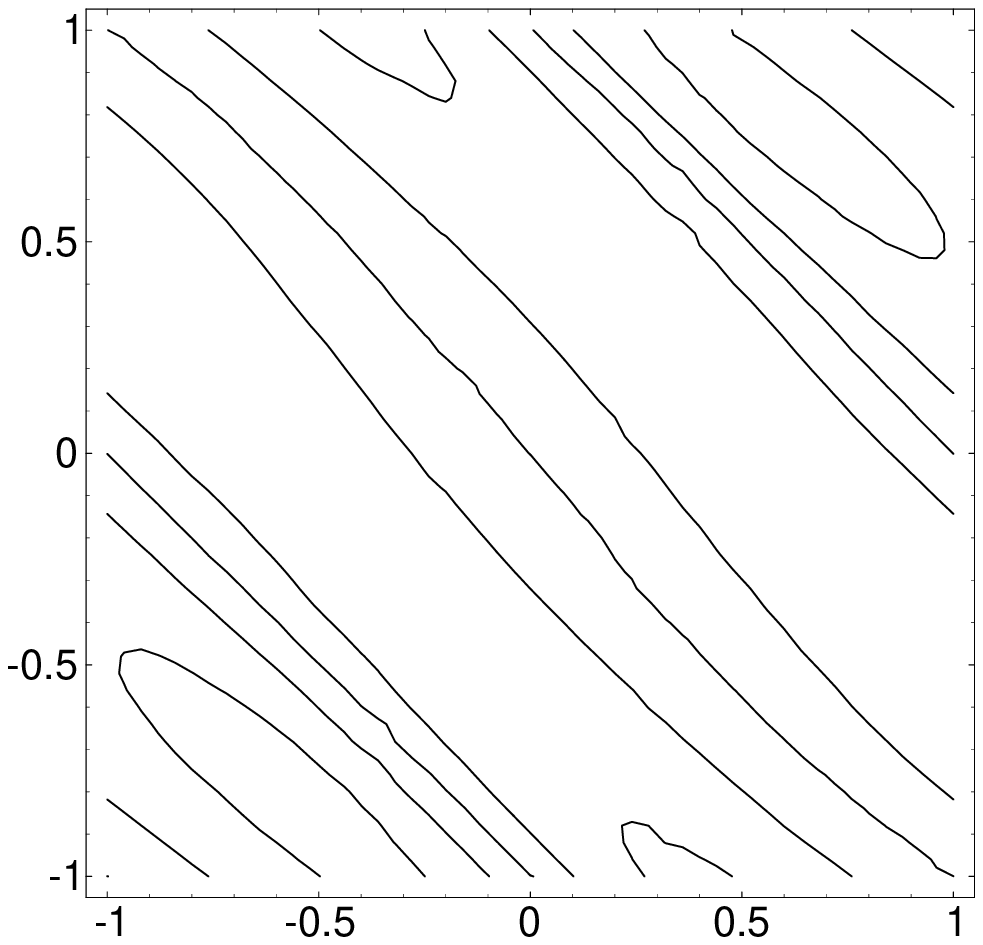}}
	\put(3.5,7.4){\fbox{${\mathcal A}_{II}$ in \% }}
	\put(6.0,-0.3){$\varphi_{\mu}[\pi]$}
	\put(0,7.4){$\varphi_{M_1}[\pi]$ }
	\put(1.1,0.8){\scriptsize 3}
	\put(1.6,1.5){\scriptsize 6}
	\put(2.,2.){\scriptsize 3}
	\put(2.25,2.43){\scriptsize 0}
	\put(2.55,2.6){\scriptsize -3}
	\put(4.4,1.2){\scriptsize -6}
	\put(3.2,3.1){\scriptsize -3}
	\put(4.0,3.7){\scriptsize 0}
	\put(4.35,4.3){\scriptsize 3}
	\put(3.1,6.4){\scriptsize 6}
	\put(5.0,5.0){\scriptsize 3}
	\put(5.3,5.2){\scriptsize 0}
	\put(5.5,5.5){\scriptsize -3}
	\put(5.8,5.8){\scriptsize -6}
	\put(6.6,6.5){\scriptsize -3}
\put(0.5,-.3){Fig.~\ref{plot3}a}
	\put(8,0){\includegraphics{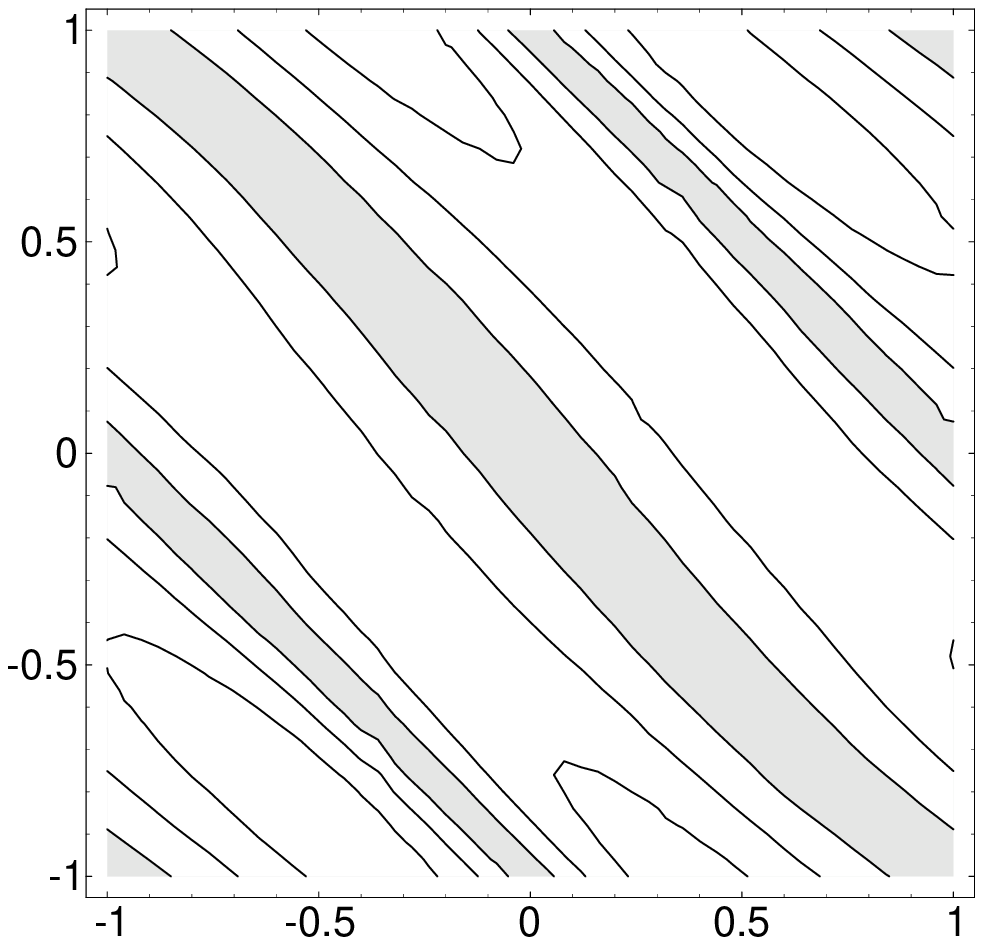}}
	\put(10.3,7.4){\fbox{$S_{II} =|{\mathcal A}_{II}| 
			\sqrt{2{\mathcal L}\cdot\sigma}$  }}
	\put(14.,-.3){$\varphi_{\mu}[\pi] $ }
	\put(8,7.4){$\varphi_{M_1}[\pi]$ }
	\put(9.1,1.36){\scriptsize 10}
	\put(9.6,1.5){\scriptsize 15}
	\put(10.,2.0){\scriptsize 10}
	\put(10.5,2.8){\scriptsize 10}
	\put(10.8,3.3){\scriptsize 10}
	\put(12.3,1.2){\scriptsize 15}
	\put(9.0,5.0){\scriptsize 15}
	\put(12.4,4.5){\scriptsize 10}
	\put(11.3,6.2){\scriptsize 15}
	\put(12.9,4.8){\scriptsize 10}
	\put(14.6,4.7){\scriptsize 10}
	\put(13.2,6.2){\scriptsize 15}
	\put(14.1,5.8){\scriptsize 15}
	\put(14.4,6.1){\scriptsize 10}
\put(8.5,-.3){Fig.~\ref{plot3}b}
\end{picture}
\vspace*{.5cm}
\caption{
	Contour lines of 
	and the asymmetry ${\mathcal A}_{II}$ (\ref{plot3}a)
	and the statistical significance $S_{II}$ (\ref{plot3}b)
	for $e^+e^-\to\tilde\chi^+_1\tilde\chi^-_1;~ 
	\tilde\chi^+_1 \to  W^+\tilde\chi^0_1 ;~W^+ \to c \bar s$,
	in the $\varphi_{\mu}$--$\varphi_{M_1}$ plane  
	for $|\mu|=350$~GeV, $M_2=400$~GeV,
	$\tan \beta=5$, $m_0=300$ GeV,
	$\sqrt{s}=800$ GeV, $(P_{e^-},P_{e^+})=(-0.8,0.6)$
	and ${\mathcal L}=500~{\rm fb}^{-1}$.
	In the gray shaded area of Fig.~\ref{plot3}b
	we have $S_{II}<5$.
	\label{plot3}}
\end{figure}

In Fig.~\ref{plot3}a we plot
the contour lines  of ${\mathcal A}_{II}$ for $|\mu|=350$~GeV and $M_2=400$~GeV
in the $\varphi_{\mu}$--$\varphi_{M_1}$ plane.
Fig.~\ref{plot3}a shows that ${\mathcal A}_{II}$ is essentially
depending on the sum $\varphi_{\mu} + \varphi_{M_1}$.
However, maximal phases of  
$\varphi_{M_1}=\pm 0.5\pi$ and $\varphi_{\mu}=\pm 0.5\pi$ 
do not  lead to the highest values of 
$|{\mathcal A}_{II}| \gsim  6\%$, which are reached for
$(\varphi_{M_1},\varphi_{\mu}) \approx (\pm0.8 \pi,\pm0.6 \pi)$.
The reason for this is that the spin-correlation terms 
$\Sigma_P^a \,^c\Sigma_{D_1}^a \,^cD_2$ in the numerator of 
${\mathcal A}_{II}$ (\ref{asymII}) are products of CP odd and
CP even factors. The CP odd (CP even) factors have a sine-like
(cosine-like) phase dependence. Therefore, the maximum of the CP
asymmetry ${\mathcal A}_{II}$ may be shifted to smaller or larger
values of the phases.
In the $\varphi_{\mu}$--$\varphi_{M_1}$ region shown 
in Fig.~\ref{plot3}a the cross section 
$\sigma=\sigma_P(e^+e^-\to\tilde\chi^+_1\tilde\chi^-_1) 
\times {\rm BR}( \tilde\chi^+_1 \to W^+\tilde\chi^0_1)
\times {\rm BR}(W^+ \to c \bar s)$,
with ${\rm BR}( \tilde\chi^+_1 \to \tilde\chi^0_1 W^+)=1$,
does not depend on $\varphi_{M_1}$ and ranges between 74 fb 
for $\varphi_{\mu}=0$ and 66 fb for $\varphi_{\mu}=\pi$.

In Fig.~\ref{plot3}b we show the contour lines of the significance
$S_{II} =|{\mathcal A}_{II}| \sqrt{2{\mathcal L}\cdot\sigma}$,
defined in Section~\ref{T odd asymmetries}.
For ${\mathcal L}= 500$ fb$^{-1}$ and for e.g. 
$(\varphi_{M_1},\varphi_{\mu}) \approx ( \pi,0.1 \pi)$
we have $S_{II} \approx 8$ and thus ${\mathcal A}_{II}$ 
could be measured 
even for small $\varphi_{\mu}$.
%

\begin{figure}[t]
\setlength{\unitlength}{1cm}
\begin{picture}(10,8)(-0.5,0)
   \put(0,0){\includegraphics{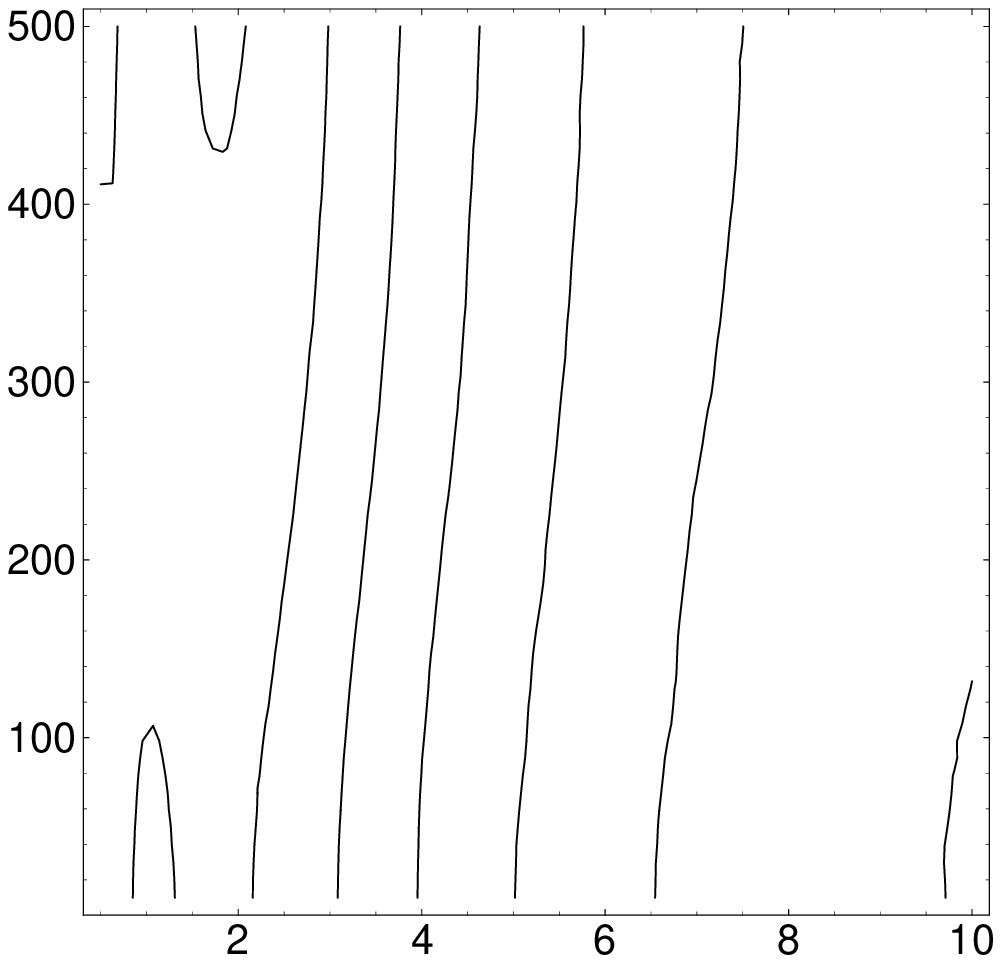}}
	\put(3.,7.4){\fbox{${\mathcal A}_{II}$ in \% }}
	\put(6.3,-0.3){$\tan\beta$}
	\put(0,7.4){$m_0[{\rm GeV}]$}
	\put(0.9,6.){\footnotesize 8}
	\put(1.05,0.9){\footnotesize 9}
	\put(1.5,6.5){\footnotesize 9}
	\put(1.9,4.){\footnotesize 8}
	\put(2.5,4.){\footnotesize 7}
	\put(3.05,4.){\footnotesize 6}
	\put(3.8,4.){\footnotesize 5}
	\put(4.8,4.){\footnotesize 4}
	\put(6.8,1.0){\footnotesize 3}
\put(0.5,-.3){Fig.~\ref{plot4}a}
\put(0.5,-.3){Fig.~\ref{plot4}a}
	\put(8,0){\includegraphics{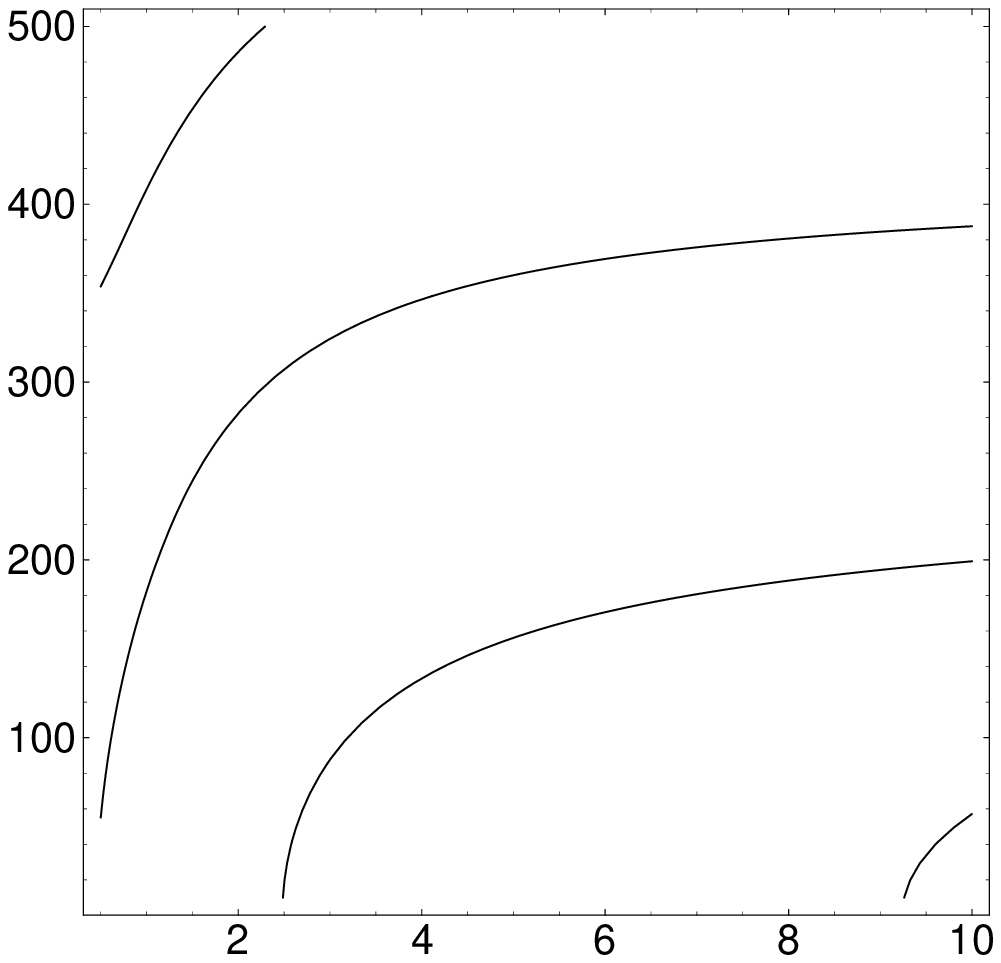}}
	\put(11.5,7.4){\fbox{$\sigma$ in fb}}
	\put(14.3,-.3){$\tan\beta$}
	\put(8,7.4){$m_0[{\rm GeV}]$}
	\put(9.5,6.0){\footnotesize 100}
	\put(11.2,4.5){\footnotesize 80}
	\put(12.3,2.3){\footnotesize 60}
	\put(14.6,0.65){\footnotesize 40}
	\put(8.5,-.3){Fig.~\ref{plot4}b}
\end{picture}
\vspace*{.5cm}
\caption{
	Contour lines of 
	the asymmetry ${\mathcal A}_{II}$ (\ref{plot4}a)
	and $\sigma=\sigma_P(e^+e^-\to\tilde\chi^+_1\tilde\chi^-_1) 
	\times {\rm BR}( \tilde\chi^+_1 \to W^+\tilde\chi^0_1)
	\times {\rm BR}(W^+ \to c \bar s)$ (\ref{plot4}b),
	in the $\tan\beta$--$m_0$ plane for 
	$(\varphi_{M_1},\varphi_{\mu})=(0.7\pi,0)$,
	$M_2=400$~GeV, $|\mu|=350$~GeV,
	$\sqrt{s}=800$ GeV and $(P_{e^-},P_{e^+})=(-0.8,0.6)$.
	\label{plot4}}
\end{figure}

In Figs.~\ref{plot4}a,b we show the 
$\tan\beta $--$m_0$ dependence of ${\mathcal A}_{II}$ and $\sigma$ 
for $(\varphi_{M_1},\varphi_{\mu})=(0.7\pi,0)$.
The asymmetry is rather insensitive to  $m_0$
and shows strong dependence on  $\tan\beta$ and
decreases with increasing $\tan\beta\gsim2$.
The production cross section
$\sigma_P(e^+e^-\to\tilde\chi^+_1\tilde\chi^-_1 )$
increases with increasing $m_0$ and decreasing $\tan\beta $.
For $m_0\lsim200$~GeV, the branching ratio 
${\rm BR}( \tilde\chi^+_1 \to W^+\tilde\chi^0_1)<1$, 
since the decay channels of $\tilde\chi^+_1$ 
into sleptons and/or sneutrinos open.
%

\begin{figure}[t]
\setlength{\unitlength}{1cm}
\begin{picture}(10,8)(-0.5,0)
	\put(-1,8){\includegraphics{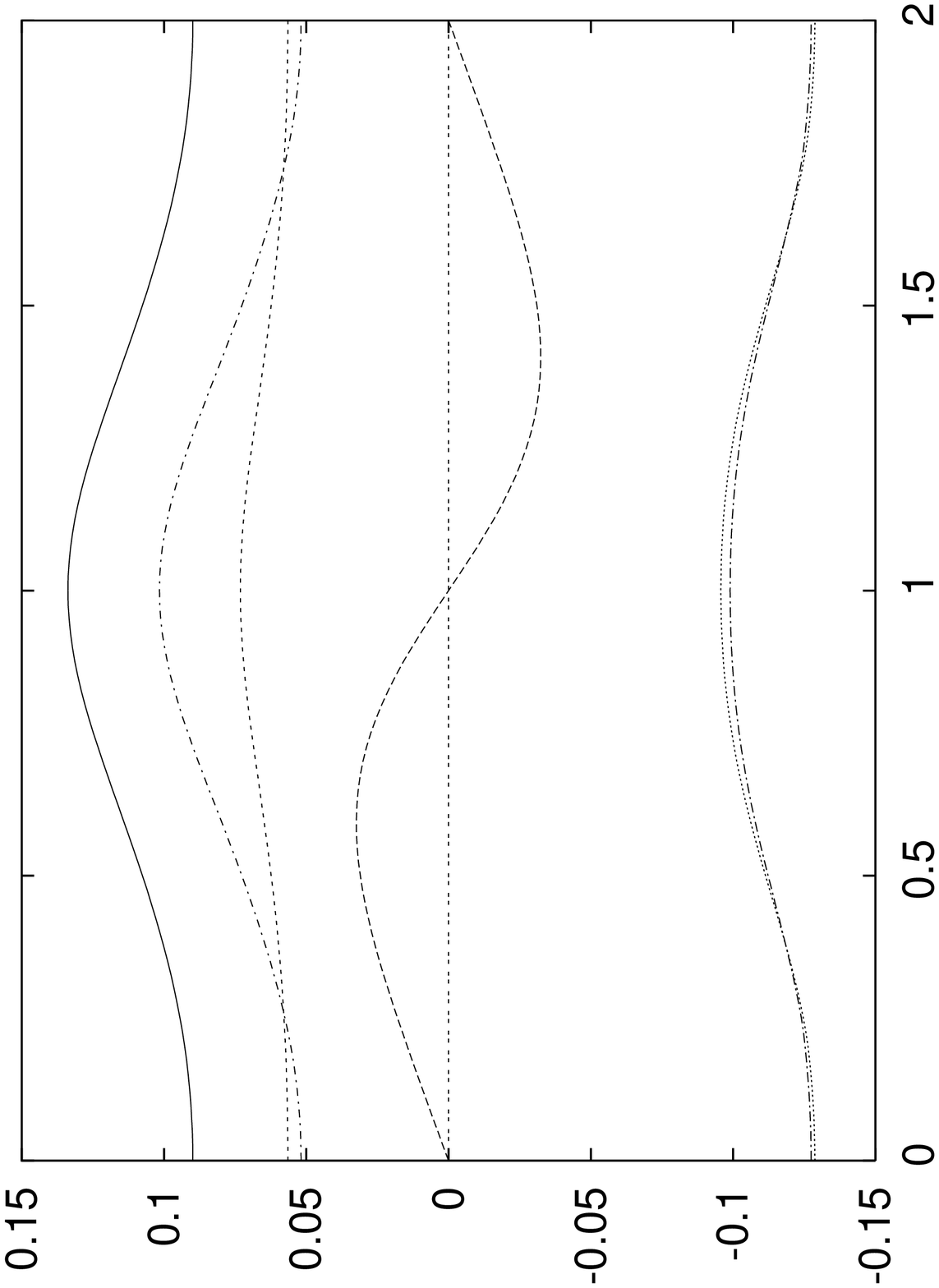}}
	\put(3.2,6.6){\footnotesize $V_1 $}
	\put(3.8,6.1){\footnotesize $ T_{13} $}
	\put(5.7,2.7){\footnotesize $ V_2$}
	\put(3.9,4.85){\footnotesize $ V_3 $}
	\put(1.3,3.2){\footnotesize $T_{12} \approx T_{23}\approx0$}
	\put(3.5,1.9){\footnotesize $ T_{11}\approx T_{22}$}
	\put(1.3,7.4){\fbox{$W$ matrix elements, $\varphi_{M_1}=\pi$}}
	\put(6.3,-0.3){$\varphi_{\mu}[\pi]$}
\put(0.5,-.3){Fig.~\ref{plot5}a}
	\put(7,8){\includegraphics{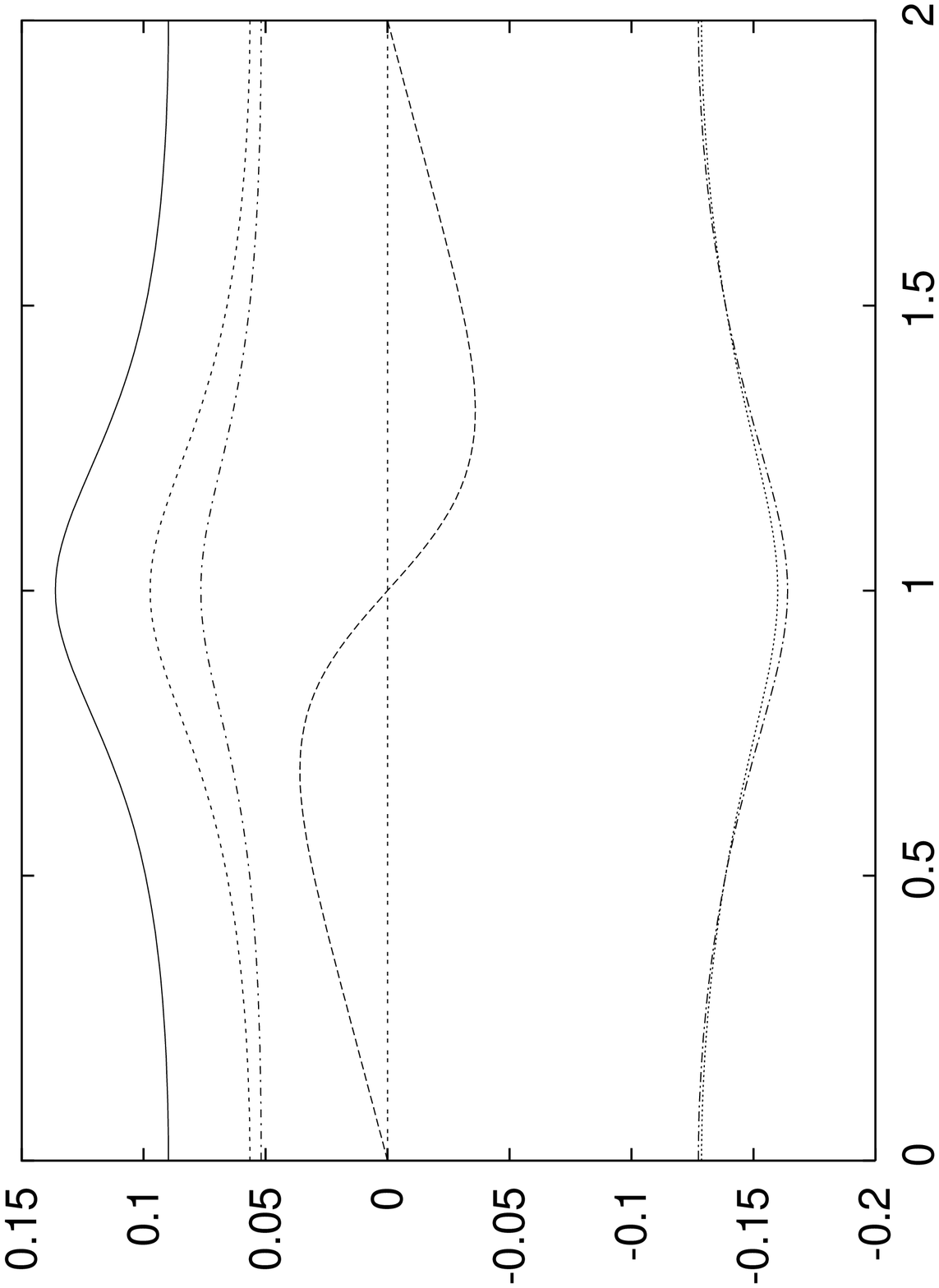}}
	\put(9.3,7.4){\fbox{$W$ matrix elements, $\varphi_{\mu}=0$ }}
	\put(14.3,-.3){$ \varphi_{M_1}[\pi]$ }
	\put(11.2,6.6){\footnotesize $V_1 $}
	\put(11.8,6.1){\footnotesize $V_3 $}
	\put(13.7,3.35){\footnotesize $ V_2$}
	\put(11.9,5.2){\footnotesize $T_{13} $}
	\put(9.4,3.75){\footnotesize $T_{12} \approx T_{23}\approx0$}
	\put(11.3,1.65){\footnotesize $ T_{11}\approx T_{22}$}
\put(8.5,-.3){Fig.~\ref{plot5}b}
\end{picture}
\vspace*{.5cm}
\caption{
   Dependence of Vector $(V_i)$ and tensor $(T_{ij})$ 
	elements of the $W^+$ density matrix  $<\rho(W^+)>$
	on $\varphi_{\mu}$  (\ref{plot5}a)
	and on $\varphi_{M_1}$ (\ref{plot5}b),
	for $e^+e^-\to\tilde\chi^+_1\tilde\chi^-_1;~ 
	\tilde\chi^+_1 \to W^+\tilde\chi^0_1 $,
	for $|\mu|=350$~GeV, $M_2=400$~GeV,
	$\tan \beta=5$, $m_0=300$ GeV,
	$\sqrt{s}=800$ GeV and $(P_{e^-},P_{e^+})=(-0.8,0.6)$.
	\label{plot5}}
\end{figure}

In Fig.~\ref{plot5}a we show the 
$\varphi_{\mu}$ dependence of the vector $(V_i)$ and tensor $(T_{ij})$ 
elements of the density matrix  $<\rho(W^+)>$
for $\varphi_{M_1}=\pi$, see (\ref{Vs}) and (\ref{Ts}).
In Fig.~\ref{plot5}b we show their dependence on $\varphi_{M_1}$
for $\varphi_{\mu}=0$.
In both figures, the element $V_2$ is CP odd,
while $T_{13}$, $T_{11}$, $T_{22}$ and 
$V_1$, $V_3$ show a CP even behavior.
As discussed in Section~\ref{Cross section},
the tensor elements $T_{11}$ and $T_{22}$ are almost 
equal and  have the same order of magnitude as $V_1$ and $V_3$, 
whereas the other elements $T_{12},|T_{23}|<10^{-5}$ are small.
In the CP conserving case
$(\varphi_{M_1},\varphi_{\mu})=(0,0)$
and $M_2=400$ GeV,  $|\mu|=350$ GeV, 
$\tan \beta=5$, $m_0=300$ GeV, 
$\sqrt{s}=800$ GeV, $(P_{e^-},P_{e^+})=(-0.8,0.6)$
the density matrix reads:
\begin{eqnarray}
	<\rho(W^+)>  =
	\left(
        \begin{array}{ccc}
			   <\rho^{--}> & <\rho^{-0}> & <\rho^{-+}> \\
			   <\rho^{0-}> & <\rho^{00}> & <\rho^{0+}> \\
			   <\rho^{+-}> & <\rho^{+0}> & <\rho^{++}>
        \end{array}
	\right) =
 	 \left(
        \begin{array}{ccc}
			   0.200 & -0.010 & -0.001\\
			  -0.010 &  0.487 &  0.137 \\
			  -0.001 &  0.137 &  0.313
        \end{array}
	\right).
\end{eqnarray}
In the CP violating case, e.g. for 
$(\varphi_{M_1},\varphi_{\mu})=(0.7\pi,0)$
and the other parameters as above, the density matrix has 
imaginary parts due to a non-vanishing $V_2$:
\begin{eqnarray}
<\rho(W^+)>  =\left(
        \begin{array}{ccc}
			  0.219        &-0.010 +0.025i & 0.002\\
			 -0.010 -0.025i& 0.405         & 0.171 + 0.025i \\
			  0.002        & 0.171 -0.025i & 0.376
        \end{array}.
	\right)
\end{eqnarray}
Imaginary parts of the density matrix are thus an 
indication of CP violation. 
%
\subsection{Production of $\tilde\chi^+_1 \, \tilde\chi^-_2$ }

\begin{figure}[t]
\setlength{\unitlength}{1cm}
\begin{picture}(10,8)(-0.5,0)
   \put(0,0){\includegraphics{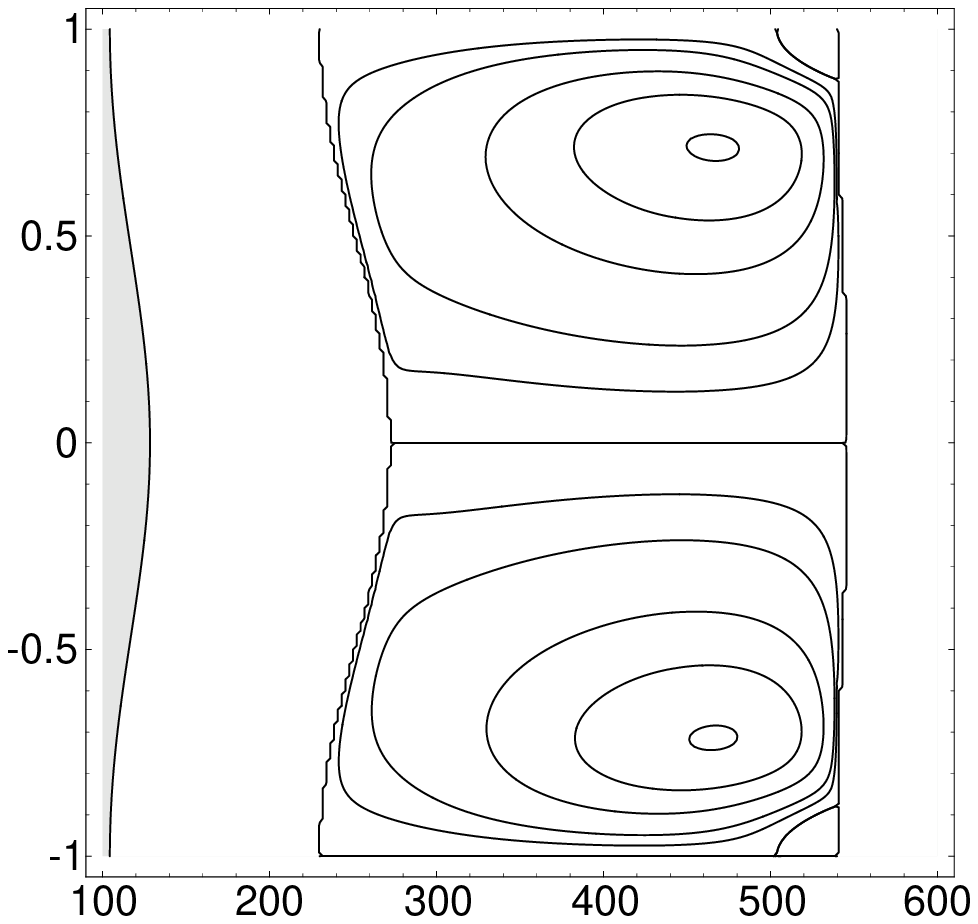}}
	\put(3.,7.4){\fbox{${\mathcal A}_{I}$ in \% }}
	\put(5.5,-0.3){$|\mu|\,[{\rm GeV}]$}
	\put(0,7.4){ $\varphi_{\mu}[\pi]$}
	\put(5.2,5.8){\scriptsize 4}
	\put(4.5,5.6){\scriptsize 3}
	\put(4.15,5.2){\scriptsize 2}
	\put(3.7,4.7){\scriptsize 1}
	\put(3.4,4.3){\scriptsize 0.5}
	\put(3.2,3.8){\scriptsize 0}
	\put(3.4,3.05){\scriptsize -0.5}
	\put(5.2,1.6){\scriptsize -4}
	\put(4.5,1.8){\scriptsize -3}
	\put(4.15,2.1){\scriptsize -2}
	\put(3.7,2.6){\scriptsize -1}
  	\put(6.6,3.7){\begin{picture}(1,1)(0,0)
			\CArc(0,0)(7,0,380)
			\Text(0,0)[c]{{\footnotesize A}}
	\end{picture}}
			\put(2.0,3.7){\begin{picture}(1,1)(0,0)
			\CArc(0,0)(7,0,380)
			\Text(0,0)[c]{{\footnotesize B}}
		\end{picture}}
\put(0.5,-.3){Fig.~\ref{plot6}a}
	\put(8,0){\includegraphics{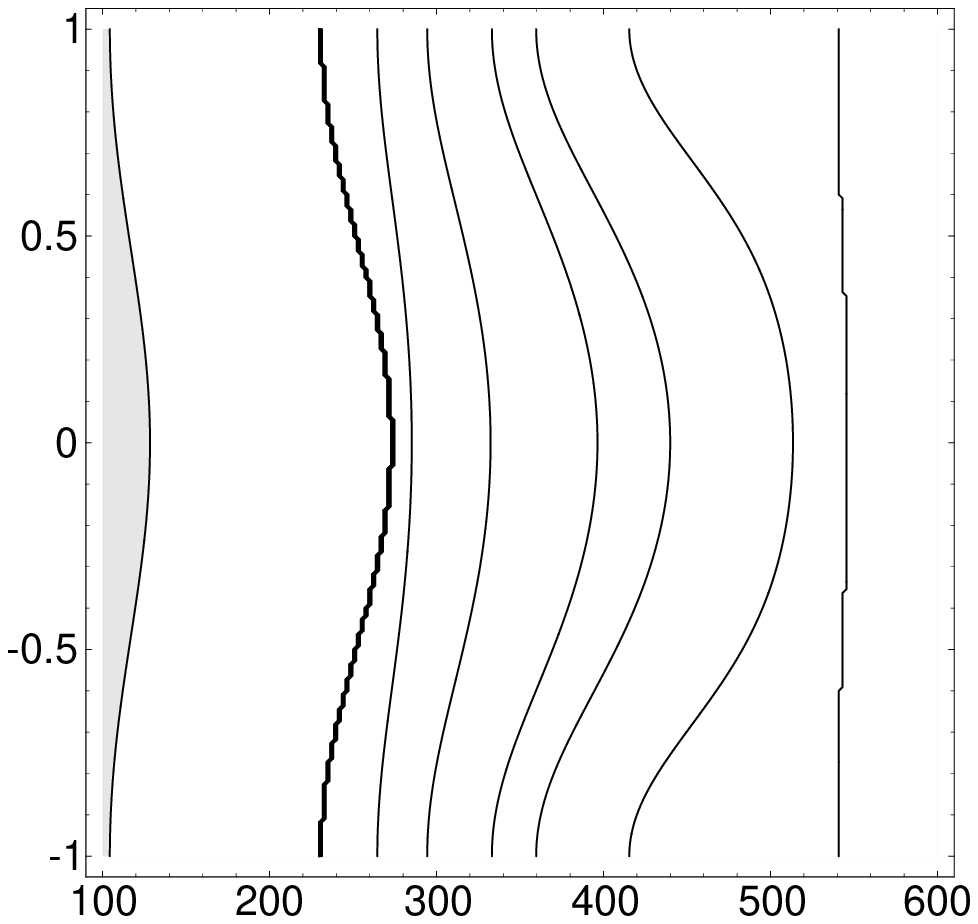}}
	\put(11.5,7.4){\fbox{$\sigma$ in fb}}
	\put(13.5,-.3){$|\mu|\,[{\rm GeV}]$}
	\put(8,7.4){ $\varphi_{\mu}[\pi]$}
	\put(11.2,3.7){\footnotesize 75}
	\put(11.7,3.7){\footnotesize 50}
	\put(12.5,3.7){\footnotesize 25}
	\put(13.0,3.7){\footnotesize 15}
	\put(13.9,3.7){\footnotesize 5}
	\put(14.6,3.7){\begin{picture}(1,1)(0,0)
			\CArc(0,0)(7,0,380)
			\Text(0,0)[c]{{\footnotesize A}}
	\end{picture}}
			\put(10.,3.7){\begin{picture}(1,1)(0,0)
			\CArc(0,0)(7,0,380)
			\Text(0,0)[c]{{\footnotesize B}}
		\end{picture}}
	\put(8.5,-.3){Fig.~\ref{plot6}b}
\end{picture}
\vspace*{.5cm}
\caption{
	Contour lines of 
	the asymmetry ${\mathcal A}_{I}$ (\ref{plot6}a)
	and $\sigma=\sigma_P(e^+e^-\to\tilde\chi^+_1\tilde\chi^-_2) 
	\times {\rm BR}( \tilde\chi^+_1 \to W^+\tilde\chi^0_1)$ 
	(\ref{plot6}b),
	in the $|\mu|$--$\varphi_{\mu}$  plane for $\varphi_{M_1}=0$, 
	$M_2=250$~GeV, $\tan \beta=5$, $m_0=300$ GeV,
	$\sqrt{s}=800$ GeV and $(P_{e^-},P_{e^+})=(-0.8,0.6)$.
		The area A is kinematically forbidden by
		$m_{\chi^+_2}+m_{\chi^-_1}>\sqrt{s}$,
		the area B by $m_{W}+m_{\chi^0_1}> m_{\chi^+_1}$.
	The gray  area is excluded by $m_{\chi_1^{\pm}}<104 $ GeV.
	\label{plot6}}
\end{figure}

For the production of an unequal pair of charginos,
$e^+e^-\to\tilde\chi^+_1\tilde\chi^-_2 $,
their polarization perpendicular to the production
plane is sensitive to the phase $\varphi_{\mu}$,
which leads to a non-vanishing asymmetry ${\mathcal A}_{I}$ (\ref{asymI}).
We will study the decay of the lighter chargino 
$\tilde\chi^+_1\to W^+\tilde\chi^0_1$.
For $|M_2|=250$~GeV and $\varphi_{M_1}=0$, we show in 
Fig.~\ref{plot6}a the  $|\mu|$--$\varphi_{\mu}$
dependence of ${\mathcal A}_{I}$, which attains values up to 4\%.
Note that ${\mathcal A}_{I}$ is not maximal for 
$\varphi_{\mu}=0.5\pi$, but is rather sensitive for phases 
in the regions $\varphi_{\mu}\in[0.7\pi,\pi]$
and $\varphi_{\mu}\in[-0.7\pi,-\pi]$. As mentioned before,
values of $\varphi_{\mu}$ close to the CP conserving points
$\varphi_{\mu}=0, \pm\pi$ are suggested by EDM analyses.
For  $\varphi_{\mu} = 0.9\pi$ and $|\mu|=350$ GeV   
the statistical significance is
$S_{I} =|{\mathcal A}_{I}|\sqrt{2{\mathcal L}\cdot\sigma} \approx 1.5$
with ${\mathcal L}=500~{\rm fb}^{-1}$.
Thus ${\mathcal A}_{I}$ could be measured 
at a confidence level larger than 68\% $(S_{I}=1)$.

In Fig.~\ref{plot6}b we show contour lines of the 
corresponding cross section 
$\sigma=\sigma_P(e^+e^-\to\tilde\chi^+_1\tilde\chi^-_2) \times
{\rm BR}( \tilde\chi^+_1 \to W^+\tilde\chi^0_1)$
in the $|\mu|$--$\varphi_{\mu}$ plane for the parameters as above.
The cross section shows a CP even behavior, 
which has been used in \cite{choi1,choigaiss,holger} 
to constrain  $\cos \varphi_{\mu}$.
In our scenario we have considered the decay of the lighter chargino 
$\tilde\chi^+_1\to W^+\tilde\chi^0_1$ 
since for our choice $m_0=300$~GeV we have  
BR$( \tilde\chi^+_1 \to W^+\tilde\chi^0_1)=1$.
For the decay of $\tilde\chi^+_2$,  
one would have to take into account also
the decays into the $Z$ boson and the lightest neutral Higgs boson, 
which would reduce BR$(\tilde\chi^+_2\to W^+\tilde\chi^0_1)\approx 0.2$.
%

\begin{figure}[t]
\setlength{\unitlength}{1cm}
\begin{picture}(10,8)(-0.5,0)
   \put(0,0){\includegraphics{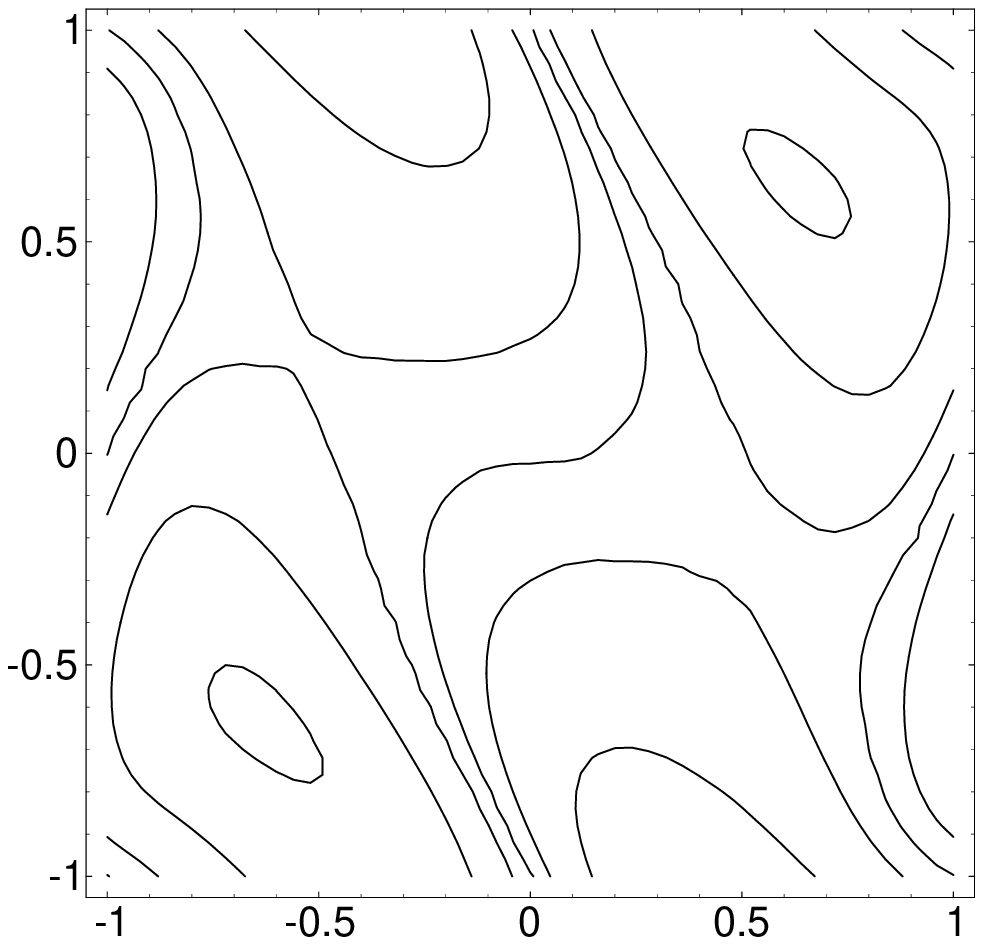}}
	\put(3.5,7.4){\fbox{${\mathcal A}_{II}$ in \% }}
	\put(6.5,-.3){$\varphi_{\mu}[\pi]$}
	\put(0.3,7.3){$ \varphi_{M_1}[\pi]$ }
	\put(1.0,1.0){\scriptsize 0}
	\put(1.2,1.5){\scriptsize 3}
	\put(1.8,1.75){\scriptsize 6.5}
	\put(1.7,4.0){\scriptsize 1}
	\put(1.6,4.8){\scriptsize 0}
	\put(1.,5.3){\scriptsize -1}
	\put(3.15,5.9){\scriptsize 3}
	\put(3.2,4.6){\scriptsize 1}
	\put(3.9,3.8){\scriptsize 0}
	\put(4.4,2.7){\scriptsize -1}
	\put(4.5,1.4){\scriptsize -3}
	\put(6.6,6.4){\scriptsize -1}
	\put(6.2,6.1){\scriptsize -3}
	\put(5.65,5.55){\scriptsize -6.5}
	\put(6.2,4.3){\scriptsize -3}
	\put(6.0,3.4){\scriptsize -1}
	\put(6.1,2.3){\scriptsize 0}
	\put(6.7,2.0){\scriptsize 1}
\put(0.5,-.3){Fig.~\ref{plot7}a}
	\put(8,0){\includegraphics{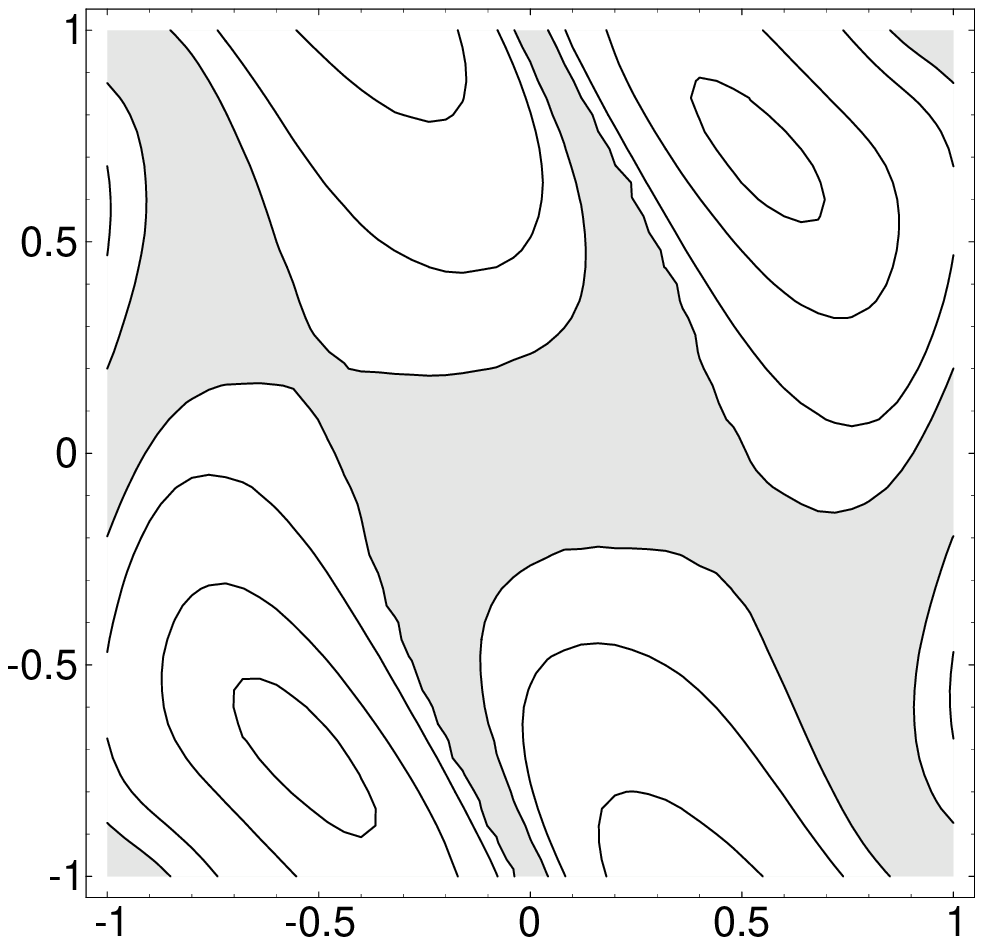}}
	\put(10.3,7.4){\fbox{$S_{II} =|{\mathcal A}_{II}| 
			\sqrt{2{\mathcal L}\cdot\sigma}$  }}
	\put(14.5,-.3){$ \varphi_{\mu}[\pi]$ }
	\put(8.5,7.4){$ \varphi_{M_1}[\pi]$ }
	\put(10.,1.8){\scriptsize 6}
	\put(9.7,2.5){\scriptsize 4}
	\put(9.5,3.3){\scriptsize 2}
	\put(9.0,5.3){\scriptsize 2}
	\put(11.2,5.2){\scriptsize 2}
	\put(11.0,6.3){\scriptsize 4}
	\put(13.3,5.9){\scriptsize 6}
	\put(14.,4.8){\scriptsize 4}
	\put(14.2,4.0){\scriptsize 2}
	\put(12.3,2.1){\scriptsize 2}
	\put(12.6,1.){\scriptsize 4}
	\put(14.75,1.9){\scriptsize 2}
\put(8.5,-.3){Fig.~\ref{plot7}b}
\end{picture}
\vspace*{.5cm}
\caption{
	Contour lines of 
	and the asymmetry ${\mathcal A}_{II}$ (\ref{plot7}a)
	and the statistical significance $S_{II}$ (\ref{plot7}b)
	for $e^+e^-\to\tilde\chi^+_1\tilde\chi^-_2;~ 
	\tilde\chi^+_1 \to W^+\tilde\chi^0_1 ;~W^+ \to c \bar s$,
	in the $\varphi_{\mu} $--$\varphi_{M_1}$ plane for 
	$|\mu|=350$~GeV, $M_2=250$~GeV, $\tan \beta=5$, 
	$m_0=300$ GeV, $\sqrt{s}=800$ GeV,
	$(P_{e^-},P_{e^+})=(-0.8,0.6)$
	and ${\mathcal L}=500~{\rm fb}^{-1}$.
	In the gray shaded area of Fig.~\ref{plot7}b
	we have $S_{II}<1$.
	\label{plot7}}
\end{figure}

The asymmetry ${\mathcal A}_{II}$  is also sensitive to the 
phase  $\varphi_{M_1}$. 
We show the $\varphi_{\mu}$--$\varphi_{M_1}$ dependence
of  ${\mathcal A}_{II}$, choosing the parameters as above, 
in Fig.~\ref{plot7}a. In Fig.~\ref{plot7}b we show the contour 
lines of the significance
$S_{II} =|{\mathcal A}_{II}| \sqrt{2{\mathcal L}\cdot\sigma}$
for ${\mathcal L}= 500$ fb$^{-1}$. For 
$(\varphi_{M_1},\varphi_{\mu}) \approx ( \pi,0.1 \pi)$
we have $S_{II} \approx 2.4$ and thus ${\mathcal A}_{II}$ could be  
accessible even for small phases by using polarized beams.
%

\section{Summary and conclusions
	\label{Summary and conclusion}}

We have proposed and analyzed CP sensitive observables in 
chargino production, $e^+e^- \to\tilde\chi^+_i  \tilde\chi^-_j$,
with subsequent two-body decay, $\tilde\chi^+_i \to W^+\chi^0_n$.
We have defined the CP asymmetry $ {\mathcal A}_{I}$ of the triple product 
${\mathbf p}_{e^-}\cdot({\mathbf p}_{\tilde\chi^+_i} \times {\mathbf p}_{W})$.
In the MSSM with complex parameters $\mu$ and $M_1$,
we have shown that ${\mathcal A}_{I}$ can reach 4\% 
and that even for $\varphi_{\mu}\approx0.9 \pi$ the asymmetry could be 
accessible in the process $e^+e^- \to\tilde\chi^+_1  \tilde\chi^-_2$.
Further we have analyzed the CP sensitive density-matrix
elements of the $W$ boson. 
The phase $ \varphi_{M_1}$ enters in the decay 
$\tilde\chi^+_i \to W^+\chi^0_n $
due to correlations of the chargino and the $W$ boson spins,
which can be probed via the hadronic decay 
$W^+ \to c \bar s$. Moreover the triple product 
${\mathbf p}_{e^-}\cdot({\mathbf p}_{c} \times {\mathbf p}_{ \bar s})$
defines the CP asymmetry $ {\mathcal A}_{II}$, 
which can be as large as 7\% for 
$\tilde\chi^+_1  \tilde\chi^-_1$ or
$\tilde\chi^+_1  \tilde\chi^-_2$ production. 
By analyzing the statistical errors of ${\mathcal A}_{I}$ and 
${\mathcal A}_{II}$ we found that 
the phases $\varphi_{\mu}$ and $ \varphi_{M_1}$
could be strongly constrained in future $e^+e^-$ collider 
experiments in the 800 GeV range with high luminosity and 
longitudinally polarized beams.

\section{Acknowledgments}

This work was supported by the 'Deutsche Forschungsgemeinschaft'
(DFG) under contract Fr 1064/5-2.
This work was also supported by the `Fonds zur
F\"orderung der wissenschaftlichen Forschung' (FWF) of Austria, project
No. P16592-N02, and by the European Community's
Human Potential Programme under contract HPRN-CT-2000-00149.

\vspace{1cm}

\begin{appendix}
	\noindent{\Large\bf Appendix}

\setcounter{equation}{0}
\renewcommand{\thesubsection}{\Alph{section}.\arabic{subsection}}
\renewcommand{\theequation}{\Alph{section}.\arabic{equation}}
\section{Coordinate frame and spin vectors
     \label{Representation of momentum and spin vectors}}
\setcounter{equation}{0}

We choose a coordinate frame in the laboratory system
such that the momentum of the chargino
$\tilde\chi ^-_j$ points in the $z$-direction
(in our definitions we follow closely \cite{gudi1}). 
The scattering  angle is 
$\theta \angle ({\mathbf p}_{e^-},{\mathbf p}_{\chi^-_j})$ and 
the azimuth $\phi$ can be chosen zero. The momenta are: 
   \begin{eqnarray}
		&& p_{e^-}^{\mu} = E_b(1,-\sin\theta,0, \cos\theta),\quad
     p_{e^+}^{\mu} = E_b(1, \sin\theta,0,-\cos\theta),\\
 &&  p_{\chi^+_i}^{\mu} = (E_{\chi^+_i},0,0,-q),\quad
     p_{\chi^-_j}^{\mu} = (E_{\chi^-_j},0,0, q),
   \end{eqnarray}
with the beam energy $E_b=\sqrt{s}/2$ and
\begin{eqnarray}
 &&   E_{\chi^+_i} =\frac{s+m_{\chi^+_i}^2-m_{\chi^-_j}^2}{2 \sqrt{s}},\quad
    E_{\chi^-_j} =\frac{s+m_{\chi^-_j}^2-m_{\chi^+_i}^2}{2 \sqrt{s}},\quad
      q =\frac{\lambda^{\frac{1}{2}}
             (s,m_{\chi^+_i}^2,m_{\chi^-_j}^2)}{2 \sqrt{s}}, 
\end{eqnarray}
where $\lambda(x,y,z) = x^2+y^2+z^2-2(xy+xz+yz)$.
For the description of the polarization of chargino 
$\tilde\chi^+_i$ we choose three spin vectors:
\begin{eqnarray}
	&&  s^{1,\,\mu}_{\tilde\chi^+_i}=(0,-1,0,0),\quad
    s^{2,\,\mu}_{\tilde\chi^+_i}=(0,0,1,0),\quad
	 s^{3,\,\mu}_{\tilde\chi^+_i}=
	 \frac{1}{m_{\tilde\chi^+_i}}(q,0,0,-E_{\tilde\chi^+_i}).
	 \label{spinvec}
\end{eqnarray} 
Together with  
$p_{\chi^+_i}^{\mu}/m_{\chi^+_i}$ they form an orthonormal set.
For the two-body decay $\tilde\chi^+_i\to W^+\tilde\chi_n^0$
the decay angle 
$\theta_{1} \angle ({\mathbf p}_{\chi^+_i},{\mathbf p}_{W})$
is constrained  by $\sin\theta^{\rm max}_{1}= q^0/q$
for $q>q^0$, where 
$q^0=\lambda^{\frac{1}{2}}(m^2_{\chi^+_i},m^2_W,m^2_{\chi^0_n})/2m_W$
is the chargino momentum if the $W$ boson is produced at rest.
In this case there are two solutions 
\begin{eqnarray}
| {\mathbf p}^{\pm}_W|= \frac{
(m^2_{\chi^+_i}+m^2_W-m^2_{\chi^0_n})q\cos\theta_{1}\pm
E_{\chi^+_i}\sqrt{\lambda(m^2_{\chi^+_i},m^2_W,m^2_{\chi^0_n})-
	 4q^2~m^2_W~(1-\cos^2\theta_{1})}}
	{2q^2 (1-\cos^2\theta_{1})+2 m^2_{\chi^+_i}}.
\end{eqnarray}
If $q^0>q$, $\theta_{1}$ is not 
constrained and there is only the physical solution 
$ |{\mathbf p}^+_W|$ left. 

The momenta in the laboratory system are
 \begin{eqnarray}
&&   p_{W}^{\pm,\,\mu} = (                        E_{W}^{\pm},
            -|{\mathbf p}_{W}^{\pm}| \sin \theta_{1} \cos \phi_{1},
             |{\mathbf p}_{W}^{\pm}| \sin \theta_{1} \sin \phi_{1},
				 -|{\mathbf p}_{W}^{\pm}| \cos \theta_{1}), \\
&& p_{\bar f}^{\, \mu} = (                        E_{\bar f},
            -|{\mathbf p}_{\bar f}| \sin \theta_{2} \cos \phi_{2},
             |{\mathbf p}_{\bar f}| \sin \theta_{2} \sin \phi_{2},
				-|{\mathbf p}_{\bar f}| \cos \theta_{2}),\\
&& E^{\mu}_{\bar f} = |{\mathbf p}_{\bar f}| = 
\frac{m_W^2}{2(E_{W}^{\pm}-|{\mathbf p}_{W}^{\pm}|\cos\theta_{D_2})},
 \end{eqnarray}
with $\theta_{2} \angle ({\mathbf p}_{\chi^+_i},{\mathbf p}_{\bar f})$
and the decay angle 
$\theta_{D_2} \angle ({\mathbf p}_{W},{\mathbf p}_{\bar f})$ given by:
 \begin{equation}
\cos\theta_{D_2}=\cos\theta_{1}\cos\theta_{2}+
\sin\theta_{1}\sin\theta_{2}\cos(\phi_{2}-\phi_{1}). 
  \end{equation}
The spin vectors $t^{c}_W$, $c=1,2,3$, of the $W$  boson
in the laboratory system are chosen as
\begin{equation}\label{defoft}
t^{1,\mu}_W=\left(0,\frac{ {\mathbf p}^2_W
\times {\mathbf p}_W^3}{|{\mathbf p}_W^2\times {\mathbf p}^3_W|}\right),\quad
t^{2,\mu}_W=\left(0,
\frac{{\mathbf p}_{e^-}\times {\mathbf p}_W}{|{\mathbf p}_{e^-}
\times {\mathbf p}_W|}\right),\quad
t^{3,\mu}_W=\frac{1}{m_W}
\left(|{\mathbf p}_W|, E_W \frac{ {\mathbf p}_W}{|{\mathbf p}_W|} \right).
  \end{equation}
The spin vectors and $p_W^{\mu}/m_W$ form an orthonormal set.
The polarization vectors 
$\varepsilon^{\lambda_k}$ 
for helicities $\lambda_k=-1,0,+1$ of the $W$ boson
are defined by:
\begin{equation}\label{circularbasis}
	\varepsilon^-={\textstyle \frac{1}{\sqrt 2}}(t^1_W-i t^2_W);
	\quad \varepsilon^0=t^3_W; \quad
	\varepsilon^+=-{\textstyle \frac{1}{\sqrt 2}}(t^1_W+i t^2_W).
\end{equation}

\section{Phase space
     \label{Phase space}}
\setcounter{equation}{0}

The Lorentz invariant phase space element for the chargino 
production (\ref{production}) and the decay 
chain (\ref{decay_1})-(\ref{decay_2}) can be decomposed
into the two-body  phase space elements:
\begin{eqnarray}
 &&d{\rm Lips}(s,p_{\chi^-_j },p_{\chi^0_n},p_{f^{'}},p_{\bar f}) =
	 \nonumber \\ 
&&\frac{1}{(2\pi)^2}~d{\rm Lips}(s,p_{\chi^+_i},p_{\chi^-_j} )
~d s_{\chi^+_i} ~\sum_{\pm}d{\rm Lips}(s_{\chi^+_i},p_{\chi^0_n},p_{W}^{\pm})
 ~d s_{W}~d{\rm Lips}(s_{W},p_{f^{'}},p_{\bar f}),\label{Lips}
 \end{eqnarray}
\begin{eqnarray}
	d{\rm Lips}(s,p_{\chi^+_i },p_{\chi^-_j })&=&
	\frac{q}{8\pi\sqrt{s}}~\sin\theta~ d\theta, \\
	d{\rm Lips}(s_{\chi^+_i},p_{\chi^0_n},p_W^{\pm})&=&
\frac{1}{2(2\pi)^2}~
\frac{|{\mathbf p}_W^{\pm}|^2}{2|E_W^{\pm}~q\cos\theta_1-
	E_{\chi^+_i}~|{\mathbf p}^{\pm}_W||}~d\Omega_1,\\
	d{\rm Lips}(s_{W},p_{f^{'}},p_{\bar f})&=&
\frac{1}{2(2\pi)^2}~\frac{|{\mathbf p}_{\bar f}|^2}{m_W^2}
	~d\Omega_2,
\end{eqnarray}
with $s_{\chi^+_i}=p^2_{\chi^+_i}$, $s_{W}=p^2_{W}$ and 
$ d\Omega_i=\sin\theta_i~ d\theta_i~ d\phi_i$.
We use the narrow width approximation for the propagators:
$\int|\Delta(\tilde\chi^+_i)|^2 $ $ d s_{\chi^+_i} = 
\frac{\pi}{m_{\chi^+_i}\Gamma_{\chi^+_i}}, ~
\int|\Delta(W)|^2 d s_{W} = 
\frac{\pi}{m_{W}\Gamma_{W}}$.
The approximation is justified for
$(\Gamma_{\chi^+_i}/m_{\chi^+_i})^2\ll1$,
which holds in our case with 
$\Gamma_{\chi^+_i}\lsim {\mathcal O}(1 {\rm GeV}) $.

\section{Spin matrices
     \label{matrices}}
\setcounter{equation}{0}
In the basis (\ref{circularbasis}) the spin matrices 
$J^{c}$ and the tensor components $J^{cd}$ are \cite{olafz}: 
 \begin{eqnarray}
  J^1=
  \left(
        \begin{array}{rrr}
			  0&\frac{1}{\sqrt{2}}&0\\
			  \frac{1}{\sqrt{2}}&0&\frac{1}{\sqrt{2}}\\
			  0&\frac{1}{\sqrt{2}}&0
        \end{array}
	  \right),
&
  J^2=
  \left(
        \begin{array}{rrr}
			  0&\frac{i}{\sqrt{2}}&0\\
			  -\frac{i}{\sqrt{2}}&0&\frac{i}{\sqrt{2}}
			  \\0&-\frac{i}{\sqrt{2}}&0
        \end{array}
	  \right),
&
  J^3=
  \left(
        \begin{array}{rrr}
         -1&0&0\\0&0&0\\0&0&1
        \end{array}
	  \right), \label{Jcircular1}\\
J^{11}=
  \left(
        \begin{array}{rrr}
			  -\frac{1}{3}&0&1\\
			  0&\frac{2}{3}&0\\
			  1&0&-\frac{1}{3}
        \end{array}
	  \right),
&
J^{22}=
\left(
	\begin{array}{rrr}
	  -\frac{1}{3}&0&-1\\
	  0&\frac{2}{3}&0\\
	  -1&0&-\frac{1}{3}
	  \end{array}
	  \right),
	  &
J^{33}=  
\left(
	\begin{array}{rrr}
	  \frac{2}{3}&0&0\\
	  0&-\frac{4}{3}&0\\
	  0&0&\frac{2}{3}
	  \end{array}
  \right),\label{Jcircular2}\\
J^{12}=
  \left(
        \begin{array}{rrr}
         0&0&i\\0&0&0\\-i&0&0
        \end{array}
	  \right),
&
J^{23}=
  \left(
        \begin{array}{rrr}
			  0&-\frac{i}{\sqrt{2}}&0\\
			  \frac{i}{\sqrt{2}}&0&\frac{i}{\sqrt{2}}
			  \\0&-\frac{i}{\sqrt{2}}&0
        \end{array}
	  \right),
&
J^{13}=
  \left(
        \begin{array}{rrr}
			  0&-\frac{1}{\sqrt{2}}&0\\
			  -\frac{1}{\sqrt{2}}&0&\frac{1}{\sqrt{2}}\\
			  0&\frac{1}{\sqrt{2}}&0
        \end{array}
	  \right). \label{Jcircular3}
\end{eqnarray}

\section{Chargino production matrices
  \label{Chargino production matrices}}
\setcounter{equation}{0}

We give the analytical formulae for
$P,\Sigma_P^1,\Sigma_P^2,\Sigma_P^3$ 
of the chargino production matrix
$\rho_P(\tilde\chi^+_i)^{\lambda_i \lambda_i'} =
  2(\delta_{\lambda_i \lambda_i'} P + 
       \sigma^{a}_{\lambda_i \lambda_i'}\Sigma_P^a)$
(\ref{rhoP}), in the laboratory system. 
Covariant expressions for these functions 
can be found in \cite{gudi1}.

\subsection{Chargino production 
     \label{Chargino production}}

The coefficient $P$ is independent of the chargino polarization. 
It can be composed into the different contributions from the 
production channels:
\begin{equation}
	P =  P(\gamma \gamma)
		+ P(\gamma Z) 
		+ P(\gamma \tilde \nu)
		+ P(Z Z)
		+ P(Z\tilde \nu)
		+ P(\tilde \nu\tilde \nu)
\end{equation}
which read
\begin{eqnarray}
P(\gamma \gamma)&=&\delta_{ij}2e^4 |\Delta (\gamma)|^2
	(c_{L} + c_{R})
	E_b^2(E_{\chi_i^+} E_{\chi_j^-}+m_{\chi_i^+} m_{\chi_j^-}+
	q^2\cos^2\theta),\\
P(\gamma Z)&=&\delta_{ij}2\frac{e^2g^2}{\cos^2 \theta_W}E_b^2
	Re\Big\{  \Delta (\gamma)\Delta (Z)^{\ast} \Big[   
	(L_e c_{L} - R_e c_{R})
	(O^{'R\ast}_{ij} - O^{'L\ast}_{ij})
	2E_bq\cos\theta
		\nonumber\\& & 
	+(L_ec_{L} + R_e c_{R})
	(O^{'L\ast}_{ij}+O^{'R\ast}_{ij})
	(E_{\chi_i^+} E_{\chi_j^-}+m_{\chi_i^+} m_{\chi_j^-}+q^2\cos^2\theta)
	\Big]\Big\},\\
P(\gamma \tilde \nu)&=&\delta_{ij}e^2g^2E_b^2
	c_{L}
	Re\Big\{ V^{\ast}_{i1}V_{j1}\Delta (\gamma)\Delta (\tilde \nu)^{\ast}
	\Big\} \times \nonumber\\& & 
	(E_{\chi_i^+} E_{\chi_j^-}+m_{\chi_i^+} m_{\chi_j^-}-2E_bq\cos\theta
	+q^2\cos^2\theta),\\
P(Z Z)&=&\frac{g^4}{\cos^4\theta_W}|\Delta (Z)|^2E_b^2\Big[
	(L_e^2 c_{L} - R_e^2  c_{R})
	(|O^{'R}_{ij}|^2-|O^{'L}_{ij}|^2)2E_bq\cos\theta
	\nonumber\\& &
	+(L_e^2c_{L} + R_e^2c_{R})
	(|O^{'L}_{ij}|^2+|O^{'R}_{ij}|^2)
	(E_{\chi_i^+} E_{\chi_j^-}+ q^2\cos^2\theta)
	\nonumber\\& &
	+(L_e^2 c_{L} + R_e^2 c_{R})
	2Re\{O^{'L}_{ij}O^{'R\ast}_{ij}\}m_{\chi_i^+} m_{\chi_j^-}
	\Big], \\
P(Z\tilde \nu)&=&\frac{g^4}{\cos^2\theta_W}
	L_e c_{L}E_b^2Re\Big\{
	V^{\ast}_{i1}V_{j1}\Delta (Z)\Delta (\tilde \nu)^{\ast} \times
	\nonumber\\& & 
	[O^{'L}_{ij}(E_{\chi_i^+} E_{\chi_j^-}-2E_bq\cos\theta+q^2\cos^2\theta) 
		+O^{'R}_{ij}m_{\chi_i^+} m_{\chi_j^-}]
	\Big\},\\
P(\tilde \nu\tilde \nu)&=& \frac{g^4}{4}c_{L}
|V_{i1}|^2|V_{j1}|^2 |\Delta (\tilde \nu)|^2
	E_b^2(E_{\chi_i^+} E_{\chi_j^-}-2E_bq\cos\theta+q^2\cos^2\theta ).
\end{eqnarray}
The propagators are defined by:
\begin{equation}
        \Delta(\gamma) = \frac{i}{p_{\gamma}^2 },\quad
		      \Delta(Z)  = \frac{i}{p_Z^2-m^2_Z+im_Z\Gamma_Z},\quad
	\Delta(\tilde \nu)  = \frac{i}{p_{\tilde \nu}^2-
			  m^2_{\tilde \nu}}. \label{eq_11}
\end{equation}
The longitudinal beam polarizations are included in the weighting factors  
\begin{equation}
c_L =(1-P_{e^-})(1+P_{e^+}), \quad c_R= (1+P_{e^-})(1-P_{e^+}).
\end{equation}

\subsection{Chargino polarization 
     \label{Chargino polarization}}
 
The coefficients $\Sigma^a_P$, 
which describe the polarization of the chargino $\tilde{\chi}^+_i$, 
decompose into:
   \begin{equation}
     \Sigma_P^a =
     \Sigma_P^a(\gamma \gamma)
   + \Sigma_P^a(\gamma Z)
   + \Sigma_P^a(\gamma \tilde \nu)
   + \Sigma_P^a(Z Z)
	+ \Sigma_P^a(Z\tilde \nu)
	+ \Sigma_P^a(\tilde \nu\tilde \nu).\label{eq_27}
\end{equation}
The contributions to the transverse $\tilde{\chi}^+_i$ polarization
in the production plane are:
\begin{eqnarray}
\Sigma_P^1(\gamma \gamma)&=&\delta_{ij}2e^4 |\Delta (\gamma)|^2
	(c_{R} - c_{L})
	E_b^2\sin\theta(m_{\chi_i^+}E_{\chi_j^-}+m_{\chi_j^-}E_{\chi_i^+}),\\
\Sigma_P^1(\gamma Z)&=&\delta_{ij}2\frac{e^2g^2}{\cos^2 \theta_W}E_b^2
	\sin\theta Re\Big\{  \Delta (\gamma)\Delta (Z)^{\ast} \Big[   
	-(L_e c_{L} + R_e c_{R})
	(O^{'R\ast}_{ij} - O^{'L\ast}_{ij})
	m_{\chi_i^+}q\cos\theta
		\nonumber\\& & 
	+(R_e c_{R} - L_e c_{L})
	(O^{'L\ast}_{ij}+O^{'R\ast}_{ij})
	(m_{\chi_i^+}E_{\chi_j^-}+m_{\chi_j^-}E_{\chi_i^+})
	\Big]\Big\},\\
\Sigma_P^1(\gamma \tilde \nu)&=&-\delta_{ij} e^2 g^2 
	c_{L}E_b^2\sin\theta
	Re\Big\{ V^{\ast}_{i1}V_{j1}\Delta (\gamma)\Delta (\tilde \nu)^{\ast}
	\Big\} \times \nonumber\\& & 
	[m_{\chi_i^+}(E_{\chi_j^-}-q\cos\theta)+m_{\chi_j^-}E_{\chi_i^+}],\\
\Sigma_P^1(Z Z)&=&\frac{g^4}{\cos^4\theta_W}|\Delta (Z)|^2E_b^2 \sin\theta\Big[
	(L_e^2 c_{L} + R_e^2 c_{R})
	(|O^{'L}_{ij}|^2-|O^{'R}_{ij}|^2)m_{\chi_i^+} q\cos\theta
	\nonumber\\& &
	+(R_e^2 c_{R} - L_e^2 c_{L})
	2Re\Big\{O^{'L}_{ij}O^{'R\ast}_{ij}\Big\}m_{\chi_j^-}E_{\chi_i^+}
	\nonumber\\& &
	+(R_e^2 c_{R} - L_e^2 c_{L})
	(|O^{'R}_{ij}|^2+|O^{'L}_{ij}|^2)m_{\chi_i^+}E_{\chi_j^-}
	\Big], \\
\Sigma_P^1(Z\tilde \nu)&=&-\frac{g^4}{\cos^2\theta_W}
	L_e c_{L} E_b^2\sin\theta Re\Big\{
	V^{\ast}_{i1}V_{j1}\Delta (Z)\Delta (\tilde \nu)^{\ast}\times
	\nonumber\\& & 
	[O^{'L}_{ij}m_{\chi_i^+}(E_{\chi_j^-}-q\cos\theta) 
		+O^{'R}_{ij}m_{\chi_j^-} E_{\chi_i^+}]
	\Big\},\\
\Sigma_P^1(\tilde \nu\tilde \nu)&=& -\frac{g^4}{4}c_{L}
	|V_{i1}|^2|V_{j1}|^2 |\Delta (\tilde \nu)|^2
	E_b^2\sin\theta m_{\chi_i^+} (E_{\chi_j^-}-q\cos\theta).
\end{eqnarray}
The contributions to the transverse $\tilde{\chi}^+_i$ polarization
perpendicular to the production plane are:
\begin{eqnarray}
\Sigma_P^2(\gamma \gamma)&=&\Sigma_P^2(\tilde \nu\tilde \nu)\;=\;0,\\
\Sigma_P^2(\gamma Z)&=&\delta_{ij}2\frac{e^2g^2}{\cos^2 \theta_W}
	(R_e c_{R} - L_e c_{L})
	Im\Big\{  \Delta (\gamma)\Delta (Z)^{\ast}    
		(O^{'R\ast}_{ij} - O^{'L\ast}_{ij})\Big\} \times
	\nonumber\\& &
	E_b^2m_{\chi_j^-}q\sin\theta,\\
\Sigma_P^2(\gamma \tilde \nu)&=&\delta_{ij} e^2 g^2 c_{L}
	Im\Big\{ V^{\ast}_{i1}V_{j1}\Delta (\gamma)\Delta (\tilde \nu)^{\ast}
	\Big\}E_b^2m_{\chi_j^-}q\sin\theta, \\
\Sigma_P^2(Z Z)&=&2\frac{g^4}{\cos^4\theta_W}|\Delta (Z)|^2
	(R_e^2 c_{R} - L_e^2 c_{L})
	Im\Big\{O^{'L}_{ij}O^{'R\ast}_{ij}\Big\}
	E_b^2m_{\chi_j^-}q\sin\theta, \\
\Sigma_P^2(Z\tilde \nu)&=&\frac{g^4}{\cos^2\theta_W}L_e c_{L} 
	Im\Big\{V^{\ast}_{i1}V_{j1}O^{'R}_{ij}
	\Delta (Z)\Delta (\tilde \nu)^{\ast}\Big\}
	E_b^2m_{\chi_j^-}q\sin\theta.
\end{eqnarray}
The contributions to the longitudinal $\tilde{\chi}^+_i$ polarization are:
\begin{eqnarray}
\Sigma_P^3(\gamma \gamma)&=&\delta_{ij}2e^4 |\Delta (\gamma)|^2
	(c_{L} - c_{R})
	E_b^2\cos\theta (q^2+  E_{\chi_i^+} E_{\chi_j^-}+m_{\chi_i^+} m_{\chi_j^-}),\\
\Sigma_P^3(\gamma Z)&=&\delta_{ij}2\frac{e^2g^2}{\cos^2 \theta_W}E_b^2
	Re\Big\{  \Delta (\gamma)\Delta (Z)^{\ast} 
	\nonumber\\& &\Big[   
	(L_e c_{L} - R_e c_{R})
	(O^{'R\ast}_{ij} + O^{'L\ast}_{ij})
	(q^2+  E_{\chi_i^+} E_{\chi_j^-}+m_{\chi_i^+} m_{\chi_j^-})\cos\theta
		\nonumber\\& & 
	+(L_e c_{L} + R_e c_{R})
	(O^{'R\ast}_{ij}-O^{'L\ast}_{ij})
	q(E_{\chi_j^-}+E_{\chi_i^+}\cos^2\theta)
	\Big]\Big\},\\
\Sigma_P^3(\gamma \tilde \nu)&=&-\delta_{ij}e^2g^2 c_{L}E_b^2
	Re\Big\{ V^{\ast}_{i1}V_{j1}\Delta (\gamma)\Delta (\tilde \nu)^{\ast}
	\Big\} \times\nonumber\\& & 
	[qE_{\chi_j^-} - (q^2+E_{\chi_i^+} E_{\chi_j^-})\cos\theta
	+qE_{\chi_i^+}\cos^2\theta- m_{\chi_i^+}  m_{\chi_j^-}\cos\theta],\\
\Sigma_P^3(Z Z)&=&\frac{g^4}{\cos^4\theta_W}|\Delta (Z)|^2E_b^2\Big[
	(L_e^2 c_{L} + R_e^2 c_{R})
	(|O^{'R}_{ij}|^2-|O^{'L}_{ij}|^2)q(E_{\chi_j^-}+E_{\chi_i^+}\cos^2\theta)
	\nonumber\\& &
	+(L_e^2 c_{L} - R_e^2 c_{R})
	2Re\Big\{O^{'L}_{ij}O^{'R\ast}_{ij}\Big\}
	m_{\chi_i^+} m_{\chi_j^-}\cos\theta
	\nonumber\\& &
	+(L_e^2 c_{L} - R_e^2c_{R})
	(|O^{'L}_{ij}|^2+|O^{'R}_{ij}|^2)
	(q^2+E_{\chi_i^+} E_{\chi_j^-})\cos\theta\Big], \\
\Sigma_P^3(Z\tilde \nu)&=&\frac{g^4}{\cos^2\theta_W} L_e c_{L} E_b^2Re\Big\{
	V^{\ast}_{i1}V_{j1}\Delta (Z)\Delta (\tilde \nu)^{\ast}
	[O^{'R}_{ij}m_{\chi_i^+} m_{\chi_j^-}\cos\theta
	\nonumber\\& & 
	-O^{'L}_{ij}(q E_{\chi_j^-}-(q^2+E_{\chi_i^+} E_{\chi_j^-})\cos\theta
	+qE_{\chi_i^+}\cos^2\theta)] 
	\Big\},\\
\Sigma_P^3(\tilde \nu\tilde \nu)&=& -\frac{g^4}{4}c_{L}
	|V_{i1}|^2|V_{j1}|^2 |\Delta (\tilde \nu)|^2 E_b^2\times
	\nonumber\\& & 
	[q E_{\chi_j^-}-(q^2+E_{\chi_i^+} E_{\chi_j^-})\cos\theta
	+qE_{\chi_i^+}\cos^2\theta].
\end{eqnarray}

\end{appendix}

\end{document}